\documentclass[10pt, journal]{IEEEtran}

\usepackage{epsfig}
\usepackage{amsmath}
\usepackage[T1]{fontenc}
\usepackage{times}
\usepackage{amsmath,bm}
\usepackage{array}
\usepackage{tabularx}
\usepackage{makecell}
\usepackage{comment}
\usepackage{lipsum}
\usepackage{multirow}
\usepackage{cuted}
\usepackage{caption}
\usepackage{subcaption}
\IEEEoverridecommandlockouts

\fontfamily{ptm}\selectfont

\begin{document}

\title{Analytical Modeling of Wi-Fi and LTE-LAA Coexistence: Throughput and Impact of Energy Detection Threshold}

\author{\IEEEauthorblockN{Morteza Mehrnoush$^*$, Vanlin Sathya$^\dag$, Sumit Roy$^*$, and Monisha Ghosh$^\dag$}\\
\IEEEauthorblockA{$^*$University of Washington, Seattle, WA-98195 \\ $^\dag$University of Chicago, Illinois-60637\\
Email: \{mortezam, sroy\}@uw.edu, \{vanlin, monisha\}@uchicago.edu.} \thanks{This work was supported by the National Science Foundation (NSF) under grant 1617153.}}

\maketitle


\begin{abstract}
With both small-cell LTE and Wi-Fi networks available as alternatives for deployment in unlicensed bands (notably 5 GHz), the investigation into their coexistence is a topic of active interest, primarily driven by industry groups. 3GPP has recently standardized LTE Licensed Assisted Access (LTE-LAA) that seeks to make LTE more co-existence friendly with Wi-Fi by incorporating similar sensing and back-off features. Nonetheless, the results presented by industry groups offer little consensus on important issues like respective network parameter settings that promote ``fair access'' as required by 3GPP. Answers to such key system deployment aspects, in turn, require credible analytical models, on which there has been little progress to date.  Accordingly, in one of the first work of its kind, we develop a new framework for estimating the throughput of Wi-Fi and LTE-LAA in coexistence scenarios via suitable modifications to the celebrated Bianchi \cite{Bianchi} model. The impact of various network parameters such as energy detection (ED) threshold on Wi-Fi and LTE-LAA coexistence is explored as a byproduct and corroborated via a National Instrument (NI) experimental testbed that validates the results for LTE-LAA access priority class 1 and 3.

\end{abstract}



\section{Introduction}

The increasing penetration of high-end handheld devices using high bandwidth applications (e.g multimedia streaming) has led to an exponential increase in mobile data traffic and a consequent bandwidth crunch. Operators have had to resort to provisioning high bandwidth end-user access via {\em small cell} LTE or 802.11 Wi-Fi networks to achieve desired per-user throughput. However, in hot-spot (very high demand) scenarios, dense deployment of such small cells inevitably leads to the need for time-sharing of the unlicensed spectrum between LTE and Wi-Fi, for example, when two (non coordinating) operators respectively deploy overlapping Wi-Fi and small-cell LTE networks. An immediate frequency band of interest for such coexistence operation is the 5 GHz UNII bands in US where a significant swath of additional unlicensed spectrum was earmarked by the FCC in 2014 \cite{FCC}. 

Wi-Fi networks have been architected for operating in unlicensed spectrum via a time-sharing mechanism among Wi-Fi nodes based on the Distributed Coordination Function (DCF), and with non-Wi-Fi networks via energy detection (ED) and dynamic frequency selection (DFS) \cite{DFS}. On the other hand, there are two specifications for unlicensed LTE operation with a view to coexistence: LTE Licensed Assisted Access (LTE-LAA) and LTE Unlicensed (LTE-U) \footnote{LTE-U is proposed for regions where Listen-Before-Talk (LBT) is not required and is promoted by the LTE-U forum \cite{LTEU}.}. LTE-LAA has been developed by 3GPP and integrates a LBT mechanism \cite{3GPP_TR, ETSILAA} - similar to Carrier-Sense-Multiple-Access with Collision Avoidance (CSMA/CA) for Wi-Fi - to enable spectrum sharing worldwide in markets where it is mandated. LTE-U employs an (adaptive) duty-cycle based approach  - denoted as Carrier Sense Adaptive Transmission (CSAT) - to adapt the ON and OFF durations for LTE channel access \cite{LTEU}. Specifically, 3GPP has sought to achieve a notion of `fair coexistence'  \cite{3GPP_TR, LTEWiFi_Fair} whereby ``LAA design should target fair coexistence with existing Wi-Fi networks to not impact Wi-Fi services more than an additional Wi-Fi network on the same carrier, with respect to throughput and latency''.

As currently specified, both LTE-LAA and LTE-U utilize carrier aggregation between a licensed and an (additional) unlicensed carrier for enhanced data throughput on the downlink (DL), and all uplink traffic is transmitted on the licensed carrier. In this work, we only focus on Wi-Fi/LTE-LAA coexistence and defer Wi-Fi/LTE-U coexistence for future work.

\begin{figure}[t]
\setlength{\belowcaptionskip}{-0.1in}
\begin{center}
\includegraphics[width=3.5in]{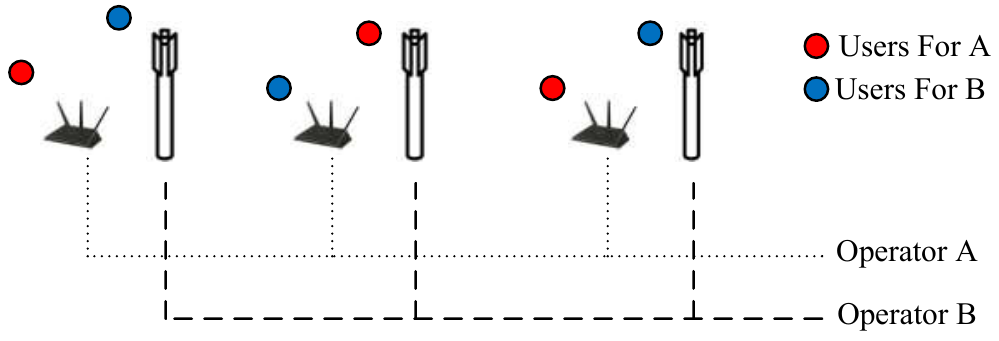}
 \caption{Wi-Fi AP (Operator A) and LTE-LAA eNB (Operator B) coexistence network scenario.}
 \label{fig: Diag}
\end{center}
\end{figure}

Despite significant efforts led by industry, there does not exist as yet a credible analytical model for investigating the coexistence mechanism proposed by 3GPP. Further, as discussed in the next section, many of the industry results based on simulations or experiments remain independently unverifiable (as can be expected - using proprietary tools or internal laboratory resources) and lack the commonly accepted analytical basis to create the necessary transparency. Consequently, results on this topic appear to be divided into two camps - one (pro-LTE) claiming that `fair' coexistence is feasible under the rules as proposed whereas the other (pro Wi-Fi)  suggesting significant negative impact and unfairness due to the presence of LTE. This has created an impasse with dueling positions based on incompatible results and no mutually acceptable pathway for crafting a methodology that builds confidence on both sides. We believe that our approach incorporating a mix of fundamental modeling backed by careful, transparent experiments will make a significant contribution towards this, and enable practical deployment of LTE-LAA LBT load based equipment (LBE) co-existing with Wi-Fi \cite{ETSILAA, 3GPP_TR} for the indoor scenario in Fig.~\ref{fig: Diag}.

The specific novel contributions of this work include: 
\begin{itemize}
\item   A new analytical model for throughput of LTE-LAA and Wi-Fi for simple coexistence scenarios consistent with 3GPP, assuming saturation;
\item  Modeling the impact of energy detection (ED) threshold and exploring its impact on the throughput of Wi-Fi and LTE-LAA;
\item Validating analytical results via experimental results using the National Instruments (NI) Labview platform.
\end{itemize}

This paper is organized as follows. Section II discusses the related research in industry and academia. Section III contains a self-contained description of the media access control (MAC) protocols of LTE-LAA and Wi-Fi. In Section IV, a new model for coexistence of Wi-Fi and LTE-LAA is developed and used to estimate throughput. Section V investigates the impact of  ED threshold on coexistence throughput.  Section VI provides detailed numerical and experimental results of Wi-Fi / LTE-LAA coexistence  and explores the impact of ED threshold. Section VII concludes the paper.

\section{Related Work}

Interest in LTE/Wi-Fi co-existence has been driven by the finalization of 3GPP Release 13 \cite{3GPP_TR}, that inspired significant industry-driven exploration of this topic.  One of the early works to explore 5 GHz LTE/Wi-Fi coexistence  \cite{LTEWiFi_Mag}  from a radio resource management perspective shows that Wi-Fi can be severely impacted by LTE transmissions implying that achieving some measure of fair coexistence of LTE and Wi-Fi needs to be carefully managed. In \cite{LTEWiFi_Per}, a simulation based performance evaluation of LTE/Wi-Fi coexistence also showed that while LTE system performance is slightly affected, Wi-Fi is significantly impacted by LTE since Wi-Fi channel access is most often blocked by LTE transmissions, causing the Wi-Fi nodes to stay in the listen mode more than 96\% of the time. 
In \cite{Google_LTEU}, the coexistence of Wi-Fi and LTE-LAA was investigated through an experimental set-up that explored the impact of the relevant carrier sensing thresholds. The LBT method in LTE-LAA employs {\em different} back-off parameter values from standard  DCF in Wi-Fi, and the consequence of this asymmetry is worthy of careful investigation\footnote{ ETSI Options A and B each specify a back-off procedure that is different from Wi-Fi.}. A second issue is the recommended sensing thresholds, for a Wi-Fi node detecting LTE transmission and vice-versa. For example, the Energy Detect (ED) threshold of -62 dBm that is used by Wi-Fi to detect any out-of-network transmission (including LTE) may not be appropriate as  LTE interference weaker than -62 dBm has been shown to be harmful to an ongoing Wi-Fi transmission. Their investigation revealed that coexistence fairness is impacted by multiple factors - the contention parameters for channel access, as well as the sensing threshold and transmission duration. \cite{LAA_GTechCableLab} also explored the relative issue of differing channelizations (i.e. channel bandwidth asymmetry) between LTE and Wi-Fi. Their results indicate that smaller bandwidth LTE-LAA transmission (e.g. $1.25 or 5$ MHz)  have a noticeable impact on Wi-Fi performance, that is dependent on where the LTE-LAA bandwidth is located relative to the Wi-Fi $20$ MHz channel. In \cite{ED_Rochman}, the authors explored the effect of ED threshold on Wi-Fi and LTE-LAA via extensive simulations and demonstrated that if both Wi-Fi and LTE employed a sensing threshold of -82 dBm to detect the other, overall throughput of both coexisting systems improved, leading to fair coexistence. 

On the other hand, Qualcomm \cite{Qualcomm} investigated the coexistence of Wi-Fi with LTE-LAA and LTE-U through simulation and showed that significant {\em throughput gain} can be achieved by aggregating LTE across licensed and unlicensed spectrum; further (and importantly), this throughput improvement does not come at the expense of degraded Wi-Fi performance and both technologies can fairly share the unlicensed spectrum. Ericsson in \cite{LTELAA_Ericsson} explored aspects of LTE-LAA system downlink (DL) operation such as dynamic frequency selection (DFS),  physical channel design, and radio resource management (RRM). An enhanced LBT approach was proposed for improving coexistence of LTE-LAA and Wi-Fi and results from a system-level simulation for 3GPP evaluation scenarios showed that fair coexistence can be achieved in both indoor and outdoor scenarios. 

In summary, industry driven research has produced mixed results: some results predict largely negative consequences for Wi-Fi with the proposed LTE-LAA coexistence mechanisms, and others that claim that fair coexistence is feasible with necessary tweaks or enhancements. There is a great need to cohere these opposing conclusions and characterize scenarios in which different conclusions are (legitimately) possible; progress on this requires a careful and transparent  approach such as ours using model-based analysis of the problem. We next summarize some of the pertinent {\em academic literature} - that while undertaking exploration of aspects of network performance in coexistence scenarios, have largely {\em not} focused on 3GPP prescriptions for 'fair' sharing.  For example, \cite{kini2016wi} explores design aspects of LBT schemes for LTE-LAA as a means of providing equal opportunity channel access in the presence of Wi-Fi. Similarly \cite{tao2015enhanced} proposed an enhanced LBT algorithm with contention window size adaptation for LTE-LAA in order to achieve fair channel access as well as Quality of Service (QoS) fairness. In \cite{han2016licensed}, authors designed the MAC protocol for LTE-LAA for fair coexistence -  the LTE transmission time is optimized for maximizing the overall normalized channel rate contributed by both LTE-LAA and Wi-Fi while protecting Wi-Fi.

The prior art on analytical models for coexistence throughput of Wi-Fi and LTE-LAA  \cite{LTEWiFi_Letter, LTEWiFi_Anal} is the closest in spirit to this contribution, based on adaptation of the Bianchi model.  In \cite{LTEWiFi_Anal}, the channel access and success probability are evaluated for the coexistence of LTE-LAA and Wi-Fi but without considering the LTE-LAA LBT implementation defined in \cite{ETSILAA, 3GPP_TR} which follows exponential back-off. Similarly, in \cite{LTEWiFi_Letter}, the coexistence of LTE-LAA and Wi-Fi throughput is evaluated for fixed contention window size (again not conformal with the 3GPP standard). In \cite{LTEWiFi_Glob}, the throughput performance of LTE-LAA and Wi-Fi is calculated for both fixed and exponential back-off; however, the packet transmission for LTE-LAA is assumed exactly the same as Wi-Fi - transmitted with $9 \mu s$ resolution slots while in reality, LTE-LAA follows the usual $0.5 ms$ transmission boundaries of LTE. In summary, a proper evaluation of coexistence of Wi-Fi and LTE-LAA LBT {\em as proposed by 3GPP} \cite{ETSILAA, 3GPP_TR} is still outstanding. 

Finally, while the imperative of much of the coexistence investigations are driven by progression to {\em fair sharing as defined by 3GPP}, we do {\em not} explore that issue in any depth in this work. This very important aspect deserves a much deeper investigation and is left for separate future work. Nonetheless, we quote a few relevant important prior art that has contributed to our understanding of the overall problem.  In \cite{LTEU_fairLeith1, LTEU_fairLeith2}, the authors derive the proportional fair rate allocation\footnote{Proportional fairness is well-known to achieve {\em airtime fairness} in rate-heterogeneous Wi-Fi networks.} for Wi-Fi/LTE-LAA (as well as Wi-Fi/LTE-U) coexistence. Also in \cite{LTEWiFi_Fair}, the fairness in the coexistence of Wi-Fi/LTE-LAA LBT based on the 3GPP criteria is investigated through a custom-built event-based system simulator. Their results suggest that LBT (and correct choice of LBT parameters) is essential to achieving proportional fairness. 

\begin{table}
\caption{Glossary of Terms}
\label{table: GlosPara}
\begin{center}
\begin{tabular}{|c|c|}
\hline
Parameter  & Definition \\
\hline
\hline
 $W_0$ & Wi-Fi minimum contention window \\
\hline
 $m$ & Wi-Fi maximum retransmission stage \\
\hline
 $W'_0$ & LTE-LAA minimum contention window \\
 \hline
 $m'$ & LTE-LAA maximum retransmission stage \\
\hline
 ACK & Acknowledgment length \\
\hline
 $\sigma$ & Wi-Fi slot time \\
\hline
 DIFS & distributed interframe space \\
\hline
 SIFS & short interframe space \\
\hline
 $N_B$ & Wi-Fi packet data portion size \\
\hline
 Psize & Wi-Fi data portion duration \\
\hline
 $T_s$ & LTE-LAA time slot for back-off \\
\hline
 $T_d$ & LTE-LAA differ time \\
 \hline
 $e_l$ & retry limit after reaching to $m'$\\
\hline
 $n_w$ & Number of Wi-Fi APs in coexistence \\
\hline
 $n_l$ & Number of LTE-LAA eNBs in coexistence \\
\hline
 $N$ & Number of Wi-Fi APs in Wi-Fi only system \\
\hline
 $P_w$ & Wi-Fi Collision Probability \\
\hline
 $\tau_w$ & a Wi-Fi station Transmission Probability \\
\hline
 $P_l$ & LTE-LAA Collision Probability \\
\hline
 $\tau_l$ & an LTE-LAA station Transmission Probability \\
\hline
 $T_D$ & transmission opportunity (TXOP) \\
\hline
 $D_{LTE}$ & Delay for next transmission \\
\hline
 $r_w$ & Wi-Fi data rate \\
\hline
 $r_l$ & LTE-LAA data rate \\
\hline
 $\delta$ & propagation delay \\
\hline
 $P_{trw}$ & Wi-Fi transmission probability \\
\hline
 $P_{sw}$ &  Wi-Fi successful transmission probability \\
\hline
 $P_{trl}$ & LTE-LAA transmission probability \\
\hline
 $P_{sl}$ &  LTE-LAA successful transmission probability \\
\hline
 $T_{sw}$ & Wi-Fi successful transmission duration \\
\hline
 $T_{cw}$ & Wi-Fi collision transmission duration \\
\hline
 $T_{sl}$ & LTE-LAA successful transmission duration \\
\hline
 $T_{cl}$ & LTE-LAA collision transmission duration \\
\hline
 $T_{cc}$ & coupled collision duration of LTE-LAA and Wi-Fi \\
\hline
 $T_{E}$ & Total average time \\
\hline
 $Tput_{w}$ & Wi-Fi throughput \\
\hline
 $Tput_{l}$ & LTE-LAA throughput \\
\hline
 $r(n)$ & sampled received signal \\
\hline
 $M$ & number of samples in energy detector \\
\hline
 $P_{dw}$ & detection probability of Wi-Fi \\
\hline
 $P_{dl}$ & detection probability of LTE-LAA \\
\hline
\end{tabular}
\end{center}
\vspace{-0.2in}
\end{table}

\begin{figure}[t]
\setlength{\belowcaptionskip}{-0.1in}
\begin{center}
\includegraphics[width=3.6in]{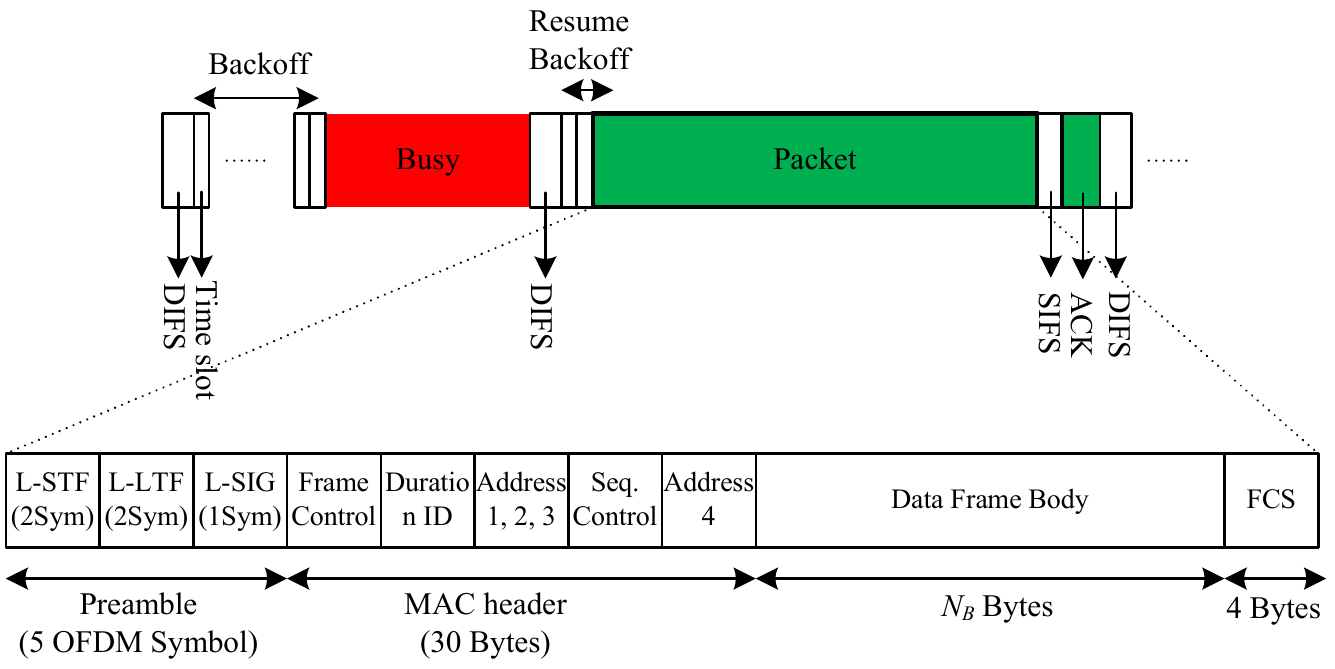}
 \caption{Wi-Fi CSMA/CA contention and frame transmission. The Wi-Fi frame structure with Preamble, MAC header, and data portion.}
 \label{fig: WiFitime}
\end{center}
\vspace{-0.2in}
\end{figure}


\section{Coexistence of LTE-LAA and Wi-Fi: MAC Protocol Mechanisms}

In this section, the MAC protocol of Wi-Fi and LTE-LAA is presented. While LTE-LAA uses LBT to mimic Wi-Fi DCF, there are some key differences which are highlighted, as these have a significant bearing on the respective channel access. 

\subsection{Wi-Fi DCF}
\label{sec: WiFiDCF}

\begin{figure}[!htb]
\setlength{\belowcaptionskip}{-0.1in}
\begin{center}
\includegraphics[width=2.5in]{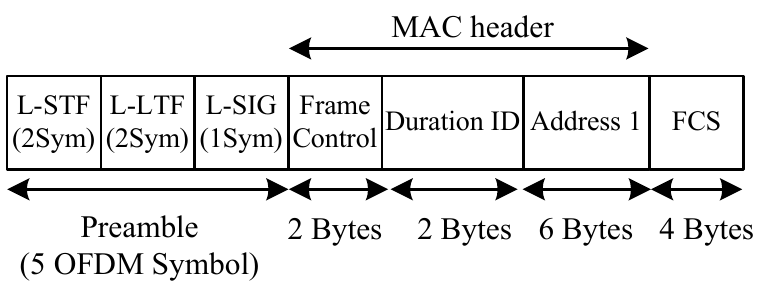}
 \caption{Wi-Fi ACK frame structure}
 \label{fig: ACKframe}
\end{center}
\vspace{-0.2in}
\end{figure}

The Wi-Fi DCF employs CSMA/CA \cite{std80211} as illustrated in Fig.~\ref{fig: WiFitime}. Each node attempting transmission must first ensure that the medium has been idle for a duration of DCF Interframe Spacing (DIFS) using the ED\footnote{The ability of Wi-Fi to detect any external or non-network interference} and carrier sensing (CS)\footnote{The ability of Wi-Fi to detect another Wi-Fi signal preamble} mechanism. When either of ED and CS is true, the Clear Channel Assessment (CCA) is indicated as busy. If the channel is idle and the station has not just completed a successful transmission, the station transmits. Otherwise, if the channel is sensed busy (either immediately or during the DIFS) or the station is contending after a successful transmission, the station persists with monitoring the channel until it is measured idle for a DIFS period, then selects a random back-off duration (counted  in units of slot time) and counts down. The back-off counter is chosen uniformly in the range  $[0, 2^i W_0 - 1]$ where the value of $i$ (the back-off stage) is initialized to 0 and $W_0$ is the {\em minimum contention window}.  Each failed transmission due to packet collision\footnote{A collision event occurs if and only if two nodes select the same back-off counter value at the end of a DIFS period.} results in incrementing the back-off stage by $1$ (binary exponential back-off or BEB) and the node counts down from the selected back-off value;  i.e. the node decrements the counter every $\sigma \, \mu s$ corresponding to a back-off slot as long as no other transmissions are detected. If during the countdown a transmission is detected, the counting is paused, and nodes continue to monitor the busy channel until it goes idle; thereafter the medium must remain idle for a further DIFS period before the back-off countdown is resumed. Once the counter hits zero, the node transmits a packet. Any node that did not complete its countdown to zero in the current round, carries over the back-off value and resumes countdown in the next round.  Once a transmission has been completed successfully, the value of $i$ is reset to 0. The maximum value of back-off stage $i$ is $m$ and it stays in $m$ stage for one more unsuccessful transmission. If the last transmission was unsuccessful, the node drops the packet and resets the back-off stage to $i=0$. If a unicast transmission is successful, the intended receiver will transmit an Acknowledgment frame (ACK) after a Short Interframe Spacing (SIFS) duration post successful reception; the ACK frame structure is shown in Fig.~\ref{fig: ACKframe} which consists of the preamble and MAC header. 

\subsection{LTE-LAA LBT}
\label{sec: LAA}

\begin{figure*}[!htb]
\setlength{\belowcaptionskip}{-0.1in}
\begin{center}
\includegraphics[width=4.5in]{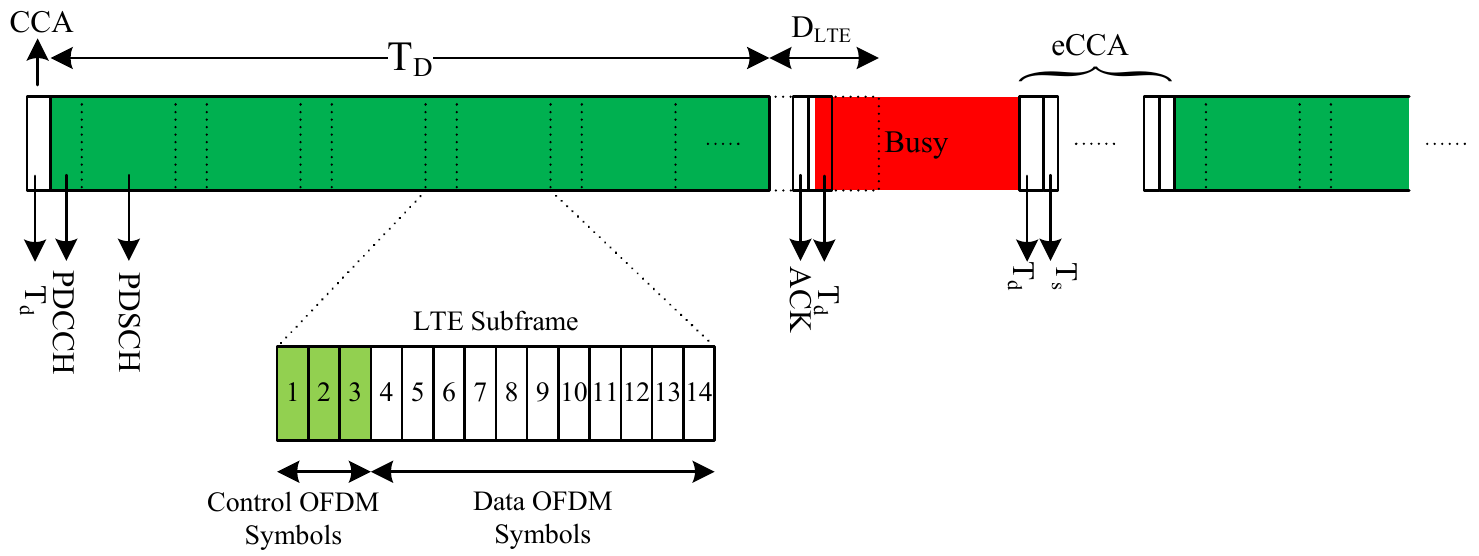}
 \caption{LTE-LAA LBT contention with CCA/eCCA and LTE subframe structure}
 \label{fig: LAAtime}
\end{center}
\vspace{-0.2in}
\end{figure*}

LTE-LAA follows a LBT approach for coexistence with Wi-Fi \cite{ETSILAA} which is similar in intent to CSMA/CA with the following key differences as illustrated in Fig.~\ref{fig: LAAtime}: \\ (a) LTE-LAA performs a CCA check using `energy detect' (CCA-ED) where it observes the channel for the {\em defer period} ($T_d$).  The $T_d$ depends on the {\em access priority class} number in Table~\ref{table: LAAclass}. There is no CS in LTE-LAA like Wi-Fi for performing preamble detection. If sensed idle and the current contention does not immediately follow a successful transmission, the LTE-LAA node starts transmission; if sensed busy, it reverts to extended CCA (eCCA) whereby it senses and defers until the channel is idle for $T_d$, and then performs the exponential back-off similar to DCF (selects a back-off counter and decrements the back-off counter every slot time $T_s=9$ $\mu s$). \\ (b) As illustrated in Table~\ref{table: LAAclass}, LTE-LAA identifies 4 channel access priority classes with different minimum and maximum contention window size. \\ (c) When a collision happens, the back-off number is selected randomly from a doubled contention window size for retransmission (i.e., $[0, 2^i W'_0 - 1]$, where $i$ is the retransmission stage for selecting the contention window size). When $i$ exceeds the maximum retransmission stage $m'$, it stays at the maximum window size for $e_l$ times ($e_l$ is the retry limit after reaching $m'$) where the $e_l$ is selected from the set of values $\{1,2,...,8\}$; then, $i$ resets to 0. \\ (d) When an LTE-LAA eNB gets access to the channel, it is allowed to transmit packets for a TXOP duration of up to 10 $ms$ when known a-priori that there is no coexistence node, otherwise up to 8 $ms$.\\ (e) The minimum resolution of data transmission in LTE-LAA is one subframe (i.e., 1 $ms$) and LTE-LAA transmits one subframe per 0.5 $ms$ slot boundary; \\ (f) After the maximum transmission time, if data is available at the LTE-LAA buffer, it should perform the eCCA for accessing the channel.

In LTE, a Resource Block (RB) is the smallest unit of radio resource which can be allocated to a UE, equal to 180 kHz bandwidth over a transmission time interval (TTI) equal to one subframe. Each RB of 180 kHz bandwidth contains 12 sub-carriers, each with 14 OFDM symbols, equaling 168 resource elements (REs). Depending upon the modulation and coding schemes (QPSK, 16-QAM, 64-QAM), each symbol or resource element in the RB carries 2, 4 or 6 bits per symbol, respectively. In an LTE-LAA system with 20 MHz bandwidth, there will be 100 RBs available. 

Each subframe consists of 14 OFDM symbols as indicated in Fig.~\ref{fig: LAAtime}, of which 1-3 are physical downlink control channel (PDCCH) symbols and the rest are physical downlink shared channel (PDSCH) data. As already mentioned, LTE-LAA eNB transmits per LTE slot boundaries, i.e. at the start of 0.5 $ms$ LTE slots for (at least) one subframe (2 LTE slots). If the eNB acquires the channel before the start of (next) LTE slot, it may need to transmit a reservation signal to reserve the channel. After the transmission period, the receiver (or receivers) transmits the ACK through the {\em licensed band} if the symbols are successfully decoded.  

\begin{table}
\caption{LTE-LAA LBT parameters per class.}
\label{table: LAAclass}
\begin{center}
\begin{tabular}{|c|c|c|c|c|c|}
\hline 
Access Priority Class \# & $T_d$ & $W'_0$ & $m'$ & TXOP \\ 
\hline
\hline
1 & 25 $\mu s$ & 4 & 1 & 2 $ms$ \\ 
 \hline
2 & 25 $\mu s$ & 8 & 1 & 3 $ms$ \\ 
 \hline
3 & 43 $\mu s$ & 16 & 2 & 8 $ms$ or 10 $ms$\\
 \hline
4 & 79 $\mu s$ & 16 & 6 & 8 $ms$ or 10 $ms$  \\
 \hline
\end{tabular}
\end{center}
\end{table}


\section{Coexistence Analysis using a modified Bianchi Model}

In this section, we first modify the 2-D Markov model in \cite{Bianchi} for analyzing the performance of the Wi-Fi DCF protocol as specified by \cite{std80211}. Then we propose a new Markov model for the  LTE-LAA LBT system. Finally, the analytical modeling for Wi-Fi and LTE-LAA coexistence (both networks use rhe same $20$ MHz channel) is used to calculate the throughput of each system when all nodes are saturated. 

\subsection{Analyzing Wi-Fi DCF using a Markov Model}

\begin{figure}[t]
\setlength{\belowcaptionskip}{-0.1in}
\begin{center}
\includegraphics[width=3.0in]{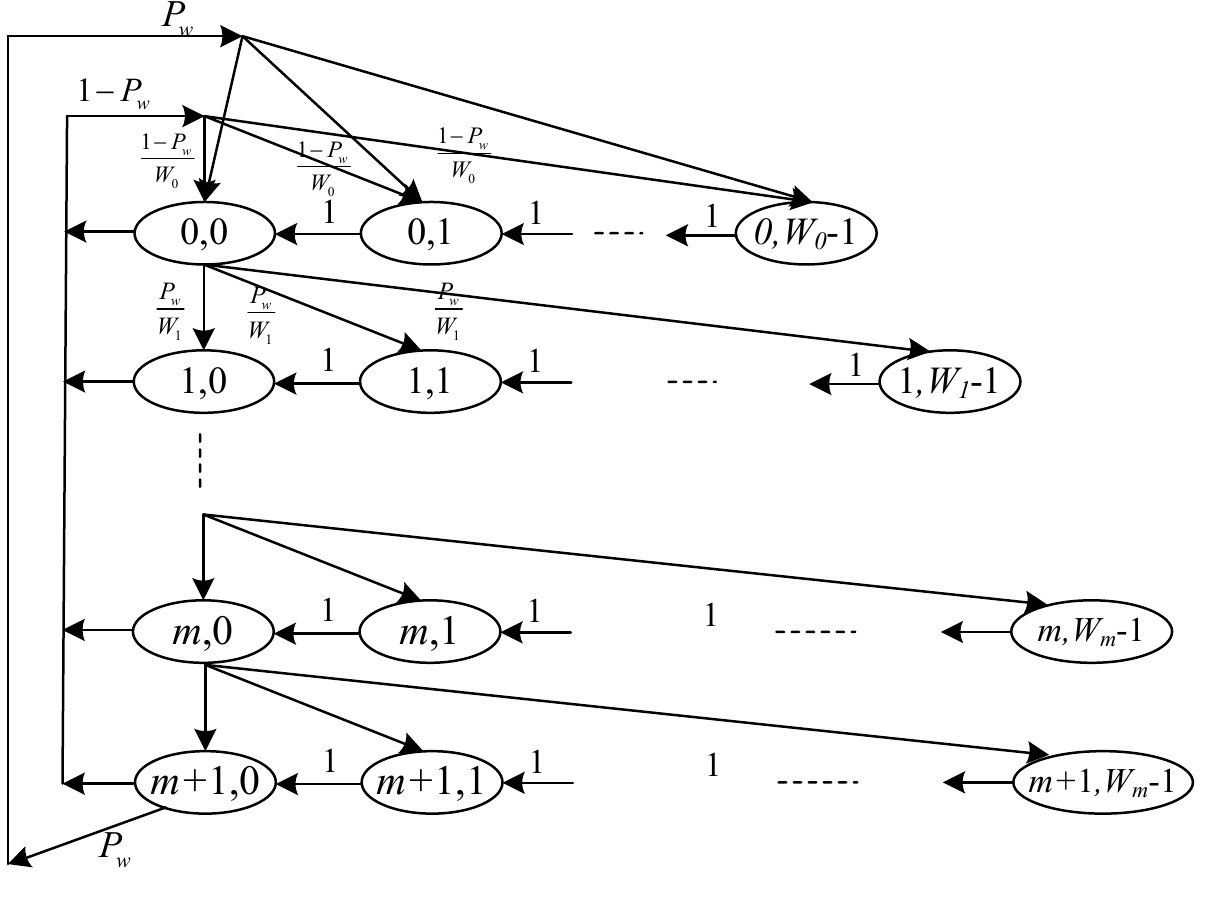}
 \caption{Markov chain model for the Wi-Fi DCF with binary exponential back-off}
 \label{fig: MarkWi-Fi}
\end{center}
\end{figure}

The 2-dimensional Markov chain model \cite{Bianchi} for Wi-Fi DCF is shown in Fig.~\ref{fig: MarkWi-Fi} for the saturated nodes. Let $\{s(t)=j,b(t)=k\}$ denote the possible states in the Markov chain, where $s(t)$ is the retransmission stage and $b(t)$ the back-off counter. The Markov chain and correspondingly one step transition probability in \cite{Bianchi} is modified based on the explanation in Section \ref{sec: WiFiDCF}. In \cite{Bianchi}, when the back-off stage reaches the maximum value $m$, it stays at $m$ forever (i.e. infinite retransmission). However, in Wi-Fi when the maximum value is reached, the back-off stage stays at $m$ for one more attempt and then resets to zero in case of an unsuccessful transmission. Therefore, the modified one step transition probabilities of the Markov chain are:
\begin{align}
 \left \{ \hspace{-0.2in} \begin{array}{c c c}
&P\{j,k|j,k+1\} = 1, ~~~ k \in (0,W_i-2) ~~~ j \in (0,m+1) \\
&P\{0,k|j,0\} = \frac{1-P_w}{W_0}, ~~~ k \in (0,W_0-1) ~~~ j \in (0,m+1) \\
&P\{j,k|j-1,0\} = \frac{P_w}{W_i}, ~~ k \in (0,W_i-1) ~~~ j \in (1,m+1) \\
&P\{0,k|m+1,0\} = \frac{P_w}{W_0}, ~~~ k \in (0,W_m-1) ~~~~~~~~~~~~~~~~ 
\end{array} \right.
\label{eq: trans1}
\end{align}
where $P_w$ is the collision probability of Wi-Fi nodes, $W_0$ is the minimum contention window size of Wi-Fi, $W_i=2^iW_0$ is the contention window size at the retransmission stage $i$, and $i=m$ is the maximum retransmission stage (i.e., $i=j$ for $j \le m$ and $i=m$ for $j > m$). The first equation in (\ref{eq: trans1}) represents the transition probability of back-off decrement; the second equation represents the transition probability after successful transmission and selecting a random back-off at stage 0 for contending for the next transmission; the third equation represents the transition probability after unsuccessful transmission in which the contention window size ($W_i$) is doubled; the last equation represents the transition probability after unsuccessful transmission in $(m+1)$th stage in which the next random back-off values should be selected from the minimum contention window size. 

Considering the stationary distribution for the Markov model as $b_{j,k}=\lim_{t\to \infty}P\{s(t)=j,b(t)=k\}, j \in (0,m+1), k \in (0,W_i-1)$, we can simplify the calculation by introducing the following variables in (\ref{eq: trans1}): 
\begin{align}
 \left \{ \begin{array}{c c}
&b_{j,0}=P_w b_{j-1,0}, ~~~~~ 0 < j \le m+1\\
&b_{j,0}=P_w^j b_{0,0}, ~~~~~~~~ 0 \le j \le m+1\\
&b_{0,0} = P_w b_{m+1,0}+(1-P_w) \sum_{j=0}^{j=m+1} b_{j,0}
\label{eq: stat1w}
\end{array} \right.,
\end{align}
the last equation implies that,
\begin{equation}
\sum_{j=0}^{j=m+1} b_{j,0}=\left(\frac{1-P_w^{m+2}}{1-P_w}\right ) b_{0,0}.
\label{eq: stat2w}
\end{equation} 
In each retransmission stage, the back-off transition probability is 
\begin{equation}
b_{j,k}=\frac{W_i-k}{W_i}b_{j,0}, ~~ 0 \le j \le m+1, ~ 0 \le k \le W_i-1.
\label{eq: stat3w}
\end{equation}

We can derive $b_{0,0}$ by the normalization condition, i.e., 
\begin{equation}
\begin{split}
&\sum_{j=0}^{m+1}\sum_{k=0}^{W_i-1}b_{j,k}=1,\\
&b_{0,0}=\frac{2}{W_0\left(\frac{(1-(2P_w)^{m+1})}{(1-2P_w)}+\frac{2^{m}\left(P_w^{m+1}-P_w^{m+2}\right)}{(1-P_w)}\right) +\frac{1-P_w^{m+2}}{1-P_w}}.
\label{eq: normw}
\end{split}
\end{equation}

Hence, the probability that a node transmits in a time slot is calculated using (\ref{eq: stat2w}) and (\ref{eq: normw}) as,
\begin{equation}
\begin{split}
&\tau_w = \sum_{j=0}^{m+1}b_{j,0}\\&=\frac{2}{W_0\left(\frac{(1-(2P_w)^{m+1})(1-P_w)+2^{m}\left(P_w^{m+1}-P_w^{m+2}\right)(1-2P_w)}{(1-2P_w)(1-P_w^{m+2})}\right) +1}.
\label{eq: tauw}
\end{split}
\end{equation}



\subsection{Analyzing LTE-LAA using a Markov Model}

\begin{figure}[t]
\setlength{\belowcaptionskip}{-0.1in}
\begin{center}
\includegraphics[width=3.0in]{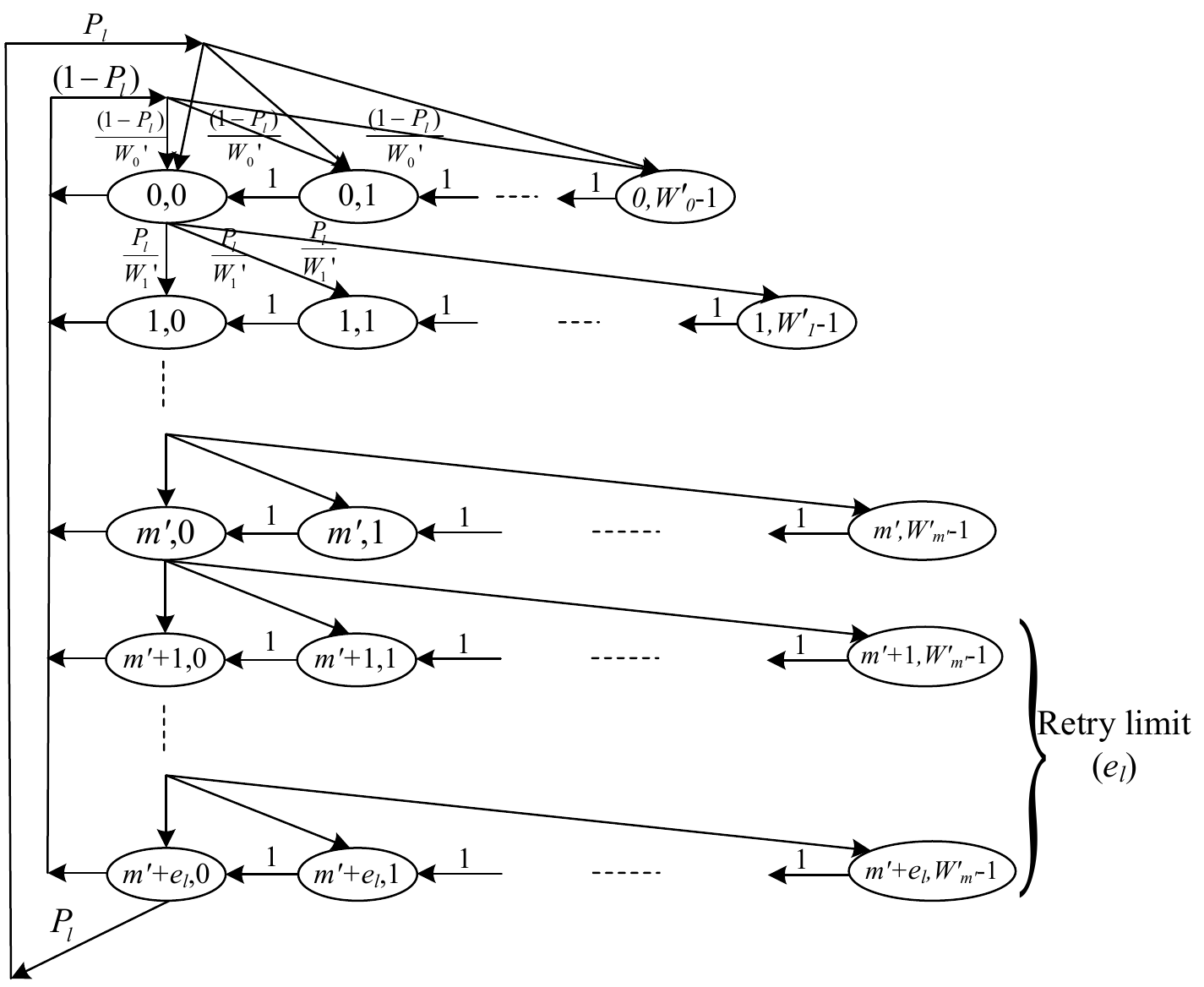}
 \caption{Markon chain model for the LTE-LAA LBT with binary exponential back-off.}
 \label{fig: MarkLTE_expo}
\end{center}
\end{figure}

In LTE-LAA, all nodes in an access priority class use the LBT mechanism for channel contention. The two-dimensional Markov chain model of LTE-LAA LBT is illustrated in Fig.~\ref{fig: MarkLTE_expo} for the saturated nodes; when the retransmission stage reaches $m'$, it stays in maximum contention window size for $e_l$ times, then resets to zero. Similarly, denoting $\{s(t)=j,b(t)=k\}$ as the states in the Markov chain, where $s(t)$ is the retransmission stage and $b(t)$ the back-off counter, the one step transition probabilities are:
\begin{align}
 \left \{ \hspace{-0.2in} \begin{array}{c c c}
&P\{j,k|j,k+1\} = 1, ~~~ k \in (0,W'_i-2) ~~~ j \in (0,m'+e_l) \\
&P\{0,k|j,0\} = \frac{1-P_l}{W'_0}, ~~~ k \in (0,W'_0-1) ~~~ j \in (0,m'+e_l) \\
&P\{j,k|j-1,0\} = \frac{P_l}{W'_i}, ~~ k \in (0,W'_i-1) ~~~ j \in (1,m'+e_l) \\
&P\{0,k|m'+e_l,0\} = \frac{P_l}{W'_0}, ~~~ k \in (0,W'_{m'}-1) ~~~~~~~~~~~~~~~
\end{array} \right. \label{eq: trans2}
\end{align}
where $P_l$ is the collision probability of LTE-LAA nodes, $W'_0$ is the minimum contention window size of LTE-LAA, $W'_i=2^iW'_0$ is the contention window size at retransmission stage $i$, and $i=m'$ is the maximum retransmission stage (i.e., $i=j$ for $j \le m'$ and $i=m'$ for $j > m'$). The first equation in (\ref{eq: trans2}) represents the transition probability of back-off decrements; the second represents the transition probability after successful transmission and selecting a random back-off at stage 0 for the contending for the next transmission; the third represents the transition probability after unsuccessful transmission in which the contention window size ($W'_i$) is doubled; the last equation represents the transition probability after unsuccessful transmission in $(m'+e_l)$th stage in which the next random back-off values reset to the minimum contention window size ($W'_0$). 

To simplify the calculation, we introduce some formulas derived from Fig.~\ref{fig: MarkLTE_expo} and (\ref{eq: trans2}) which relates the states $b_{j,k}$ to $b_{0,0}$: 
\begin{align}
 \left \{ \begin{array}{c c}
&b_{j,0}=P_l b_{j-1,0}, ~~~~~~0 < j \le m'+e_l\\
&b_{j,0}=P_l^j b_{0,0}, ~~~~~~~~0 \le j \le m'+e_l\\
&b_{0,0} = P_l b_{m'+e_l,0}+(1-P_l) \sum_{j=0}^{j=m'+e_l} b_{j,0} 
\label{eq: stat1}
\end{array} \right.,
\end{align}
the last equation is rewritten as,
\begin{equation}
\sum_{j=0}^{j=m'+e_l} b_{j,0}=\left(\frac{1-P_l^{m'+e_l+1}}{1-P_l}\right ) b_{0,0}.
\label{eq: stat2}
\end{equation} 

For each retransmission stage, the back-off transition probability is 
\begin{equation}
b_{j,k}=\frac{W'_i-k}{W'_i}b_{j,0}, ~~ 0 \le j \le m'+e_l, ~ 0 \le k \le W'_i-1
\label{eq: stat3}
\end{equation}

The $b_{0,0}$ can be derived by imposing the normalization condition as
\begin{equation}
\begin{split}
\sum_{j=0}^{m'+e_l}\sum_{k=0}^{W'_i-1}b_{j,k}=
\sum_{j=0}^{m'}\sum_{k=0}^{W'_j-1}b_{j,k}+
\sum_{j=m'+1}^{m'+e_l}\sum_{k=0}^{W'_{m'}-1}b_{j,k}=1,
\label{eq: norml}
\end{split}
\end{equation}
where $b_{0,0}$ is derived as in (\ref{eq: b00}).
\begin{floatEqTop}
\begin{align}
b_{0,0} = \frac{2}{W_0'\left(\frac{1-(2P_l)^{m'+1}}{1-2P_l}+2^{m'}\frac{P_l^{m'+1}-P_l^{m'+e_l+1}}{1-P_l}\right) +\frac{1-P_l^{m'+e_l+1}}{1-P_l}}.
\label{eq: b00}
\end{align}
\end{floatEqTop}

Hence, the probability that a node transmits in a time slot is calculated as,
\begin{equation}
\begin{split}
&\tau_l = \sum_{j=0}^{m'+e_l}b_{j,0}=\\
&\frac{2}{W_0'\left(\frac{(1-P_l)(1-(2P_l)^{m'+1})}{(1-2P_l)(1-P_l^{m'+e_l+1})}+2^{m'}\frac{P_l^{m'+1}-P_l^{m'+e_l+1}}{1-P_l^{m'+e_l+1}}\right)+1}.
\label{eq: taul}
\end{split}
\end{equation}



\subsection{Throughput  of Wi-Fi and LTE-LAA in Coexistence}

We assume that there are $n_w$ Wi-Fi APs and $n_l$ LTE-LAA eNBs which are co-channel and co-located, each with full buffer. To be consistent with 3GPP, we consider only downlink (DL) transmission (one client per AP/eNB), implying that the contention is between only the APs and eNBs. We denote $\tau_w$ ($\tau_l$) to be the access probability of a Wi-Fi (LTE-LAA) node in each time slot. Thus, for a network with $n_w$ Wi-Fi APs and $n_l$ LTE-LAA eNBs, the collision probability of a Wi-Fi AP with at least one of the other remaining ($n_w-1$ Wi-Fi and $n_l$ LTE-LAA) stations is given by,

\begin{equation}
P_w =1-(1-\tau_w)^{n_w-1} (1-\tau_l)^{n_l},
\label{eq: Pw}
\end{equation}
where $P_w$ is now coupled to both Wi-Fi and LTE-LAA nodes via $\tau_w$ and $\tau_l$.  Similarly, the collision probability for an LTE-LAA eNB with at least one of the other remaining ($n_w$ Wi-Fi and $n_l-1$ LTE-LAA) stations is,

\begin{equation}
P_l =1-(1-\tau_l)^{n_l-1} (1-\tau_w)^{n_w},
\label{eq: Pl}
\end{equation}
where $P_l$ depends on both Wi-Fi and LTE-LAA via $\tau_w$ and $\tau_l$. In order to compute the $P_w$, $P_l$, $\tau_w$, and $\tau_l$ for the coexistence of Wi-Fi and LTE-LAA, we need to jointly solve (\ref{eq: tauw}), (\ref{eq: taul}), (\ref{eq: Pw}), and (\ref{eq: Pl}). 

The transmission probability of Wi-Fi, which depends on the contention parameters of both Wi-Fi and LTE-LAA through $\tau_w$, is the probability that at least one of the $n_w$ stations transmit a packet during a time slot:
\begin{equation}
P_{trw} =1-(1-\tau_w)^{n_w},
\label{eq: Ptrw}
\end{equation}
and similarly the transmission probability of  LTE-LAA is:
\begin{equation}
P_{trl} =1-(1-\tau_l)^{n_l}.
\label{eq: Ptrl}
\end{equation}

The successful transmission of a  Wi-Fi node is the event that exactly one of the $n_w$ stations makes a transmission attempt given that at least one of the Wi-Fi nodes transmit:
\begin{equation}
P_{sw} =\frac{n_w \tau_w (1-\tau_w)^{n_w-1}}{P_{trw}},
\label{eq: Psw}
\end{equation}
Similarly the successful transmission probability of LTE-LAA is calculated as:
\begin{equation}
P_{sl} =\frac{n_l \tau_l (1-\tau_l)^{n_l-1}}{P_{trl}}.
\label{eq: Psl}
\end{equation}

Through the interdependence of Wi-Fi and LTE-LAA in the access probability of Wi-Fi and LTE-LAA, the transmission probability of Wi-Fi and LTE-LAA as well as the successful transmission probability are affected. 
To compute average throughput, we need the average time durations for a successful transmission and a collision event, respectively, given by: 
\begin{equation}
\begin{split}
T_{sw} = &\text{MACH}+\text{PhyH}+\text{Psize}+\text{SIFS}+\delta+\\
    &\text{ACK}+\text{DIFS}+\delta \\
T_{cw} = &\text{MACH}+\text{PhyH}+\text{Psize}+\text{DIFS}+\delta
\end{split},
\end{equation}
where the values of the parameters are listed as required for calculation (the parameters  presented in terms of number of bits, are converted to time based on the channel data rate provided in the numerical results section). 

The average time duration of successful transmission and collision event for LTE-LAA are 
\begin{equation}
\begin{split}
T_{sl} &= T_D+D_{LTE}\\
T_{cl} &= T_D+D_{LTE},
\end{split}
\end{equation}
where the $T_D$ is the TXOP of LTE-LAA - which could be upto 10 $ms$ for access priority class 3 and 4 \cite{ETSILAA}. $D_{LTE}$ is the delay for the next transmission which is one LTE slot (0.5 $ms$). 
After transmission for TXOP, the transmitter waits for the ACK and then resumes contention for the channel for the next transmission. If an LTE eNB wins channel contention {\em before} the start of the next LTE slot, it transmits a reservation signal to reserve the channel until the end of the current LTE slot to start packet transmission. This means that LTE-LAA contends in the $\sigma=9 \mu s$ time slot, similar to Wi-Fi, but after accessing the channel it begins its data transmission based on the LTE slot.

The throughput of Wi-Fi is calculated as:
\begin{equation}
Tput_{w} =\frac{P_{trw} P_{sw} (1-P_{trl}) \text{Psize}}{ T_E }r_w,
\label{eq: tputw}
\end{equation}
where $P_{trw} P_{sw} (1-P_{trl})$ is the probability that Wi-Fi transmits a packet successfully in one Wi-Fi slot time, and $r_w$ is the Wi-Fi physical layer data rate. $T_E$ is the average time of all possible events given by,
\begin{equation}
\begin{split}
T_{E} &=(1-P_{trw})(1-P_{trl}) \sigma+P_{trw}P_{sw}(1-P_{trl})T_{sw}\\&+P_{trl}P_{sl}(1-P_{trw})T_{sl}+P_{trw}(1-P_{sw})(1-P_{trl})T_{cw}\\&+P_{trl}(1-P_{sl})(1-P_{trw})T_{cl}+
(P_{trw}P_{sw}P_{trl}P_{sl} \\&+ P_{trw}P_{sw}P_{trl}(1-P_{sl})+ P_{trw}(1-P_{sw})P_{trl}P_{sl}  \\&+ P_{trw}(1-P_{sw})P_{trl}(1-P_{sl}))T_{cc},
\end{split}
\label{eq: TE}
\end{equation}
where $T_{cc}$ is the average time of the collision between Wi-Fi APs and LTE-LAA eNBs, determined by the larger value between $T_{cw}$ and $T_{cl}$.  

Similarly the throughput of LTE-LAA is calculated as
\begin{equation}
Tput_{l} =\frac{P_{trl} P_{sl} (1-P_{trw}) \frac{13}{14} T_D}{ T_E }r_l,
\label{eq: tputl}
\end{equation}
where $\frac{13}{14} T_D$ is the fraction of the TXOP in which the data is transmitted, i.e. 1 PDCCH symbol in a subframe with 14 OFDM symbols is considered, and $r_l$ is the LTE-LAA data rate.


\section{Impact of Energy Detect (ED) Threshold on Wi-Fi and LTE-LAA Coexistence}

We next investigate the effect of changing the ED threshold on the throughput performance of Wi-Fi and LTE-LAA in a coexistence network. In contending for channel access, Wi-Fi performs preamble based CS\footnote{Preamble based CS is near perfect because of its low carrier sensing threshold of $-82$ dBm} for detecting other Wi-Fi stations and ED for detecting external interference. In contrast, LTE-LAA uses CCA-ED detection for detecting both in and out of network transmissions. The ED threshold in generic Wi-Fi system is $-62$ dBm and $-72$ dBm in LTE-LAA. The accuracy of preamble detection for low threshold is very high, but cross-network  ED based detection is imperfect at low threshold values. Varying ED thresholds of Wi-Fi and LTE-LAA networks thus leads to varying hidden node problems (whereby additional packet drops occur due to the resulting cross-network interference) that impact the respective networks differently and lead to divergent network throughputs.

\subsection{Probability of Detection in Energy Detector}

Considering a sampling rate of 50 $ns$ in a 20 MHz Wi-Fi channel, the received signal in the presence of interference ($\mathcal{H}_1$) and no interference ($\mathcal{H}_0$) is:

\begin{equation}
\begin{split}
&    \mathcal{H}_0:(\text{W/O Interference}) ~ r(n) = w(n)  \\
&    \mathcal{H}_1:(\text{W Interference}) ~ r(n) = x_s(n)*h(n)+w(n),
\label{eq: receive}
\end{split}
\end{equation}
where $r(n)$ is the received signal, $w(n)$ is the AWGN noise, $x_s(n)$ is the modulated interference signal, and $h(n)$ is the channel impulse response (assume normalized channel, i.e. $\sum_{n}|h(n)|^2=1$). The test statistic for the energy detector (ED) is:
\begin{equation}
\mathcal{\epsilon}=\frac{1}{M}\sum_{i=1}^{M}|r(i)|^2,
\label{eq: Ttest}
\end{equation}
where $M$ is the length of received sample sequence for test statistics. For Wi-Fi DIFS duration of 34 $\mu s$, $M = 680$, the probability of detection is calculated as \cite{Energy2}:
\begin{equation}
P_d = P(\epsilon>\eta)  = Q \left ( \frac{\eta-(\sigma_n^2+\sigma_x^2)}{\frac{2}{M}(\sigma_x^2+\sigma_n^2)^2} \right),
\label{eq: Pd}
\end{equation}
where $\eta$ is the ED threshold, $\sigma_x^2$ is the signal power and $\sigma_n^2$ the noise power. Given ED threshold and number of samples, the detection probability can be calculated. 

\subsection{Modifying Analytical Model to Capture ED Threshold}

In order to incorporate the impact of ED threshold on cross network detection, we introduce $P_{dw}$ as the cross network ED detection probability of the Wi-Fi AP and $P_{dl}$ as the cross network ED detection probability of the LTE-LAA eNB. 
To incorporate the ED detection probability of Wi-Fi, we first begin by rewriting the  Wi-Fi collision probability (\ref{eq: Pw}) as
\begin{equation}
P_w =(1-(1-\tau_l)^{n_l})(1-\tau_w)^{n_w-1}+1-(1-\tau_w)^{n_w-1},
\label{eq: Pw2}
\end{equation}
where $(1-(1-\tau_l)^{n_l})$ is the probability that at least one of the LTE-LAA eNBs transmit, i.e it is the probability that the LTE-LAA is active (or ON). To capture the imperfect detection of Wi-Fi nodes ($P_{dw}$), which results in additional `physical layer collision' and consequent packet loss, we multiply the $(1-(1-\tau_l)^{n_l})$ term in eq. (\ref{eq: Pw2}) by the $P_{dw}$ as,
\begin{equation}
P_w =\left [ (1-(1-\tau_l)^{n_l})P_{dw}\right ] (1-\tau_w)^{n_w-1}+1-(1-\tau_w)^{n_w-1}.
\label{eq: Pw3}
\end{equation}


Similarly the collision probability of LTE-LAA based on eq. (\ref{eq: Pl}) can be recalculated considering the LTE-LAA detection probability ($P_{dl}$) as:

\begin{equation}
P_l =\left [ (1-(1-\tau_w)^{n_w})P_{dl}\right ] (1-\tau_l)^{n_l-1}+1-(1-\tau_l)^{n_l-1}.
\label{eq: Pl2}
\end{equation}

Using these modified equations, the analytical throughput in the previous section can be re-calculated as a function of these detection probability. 


\section{Experimental and Numerical Results}

In this section, we first present experimental results obtained using the National Instruments (NI) Labview platform for validation of our analytical model. Further, we extend the numerical results to capture the effect of different parameters in different LTE-LAA access priority classes and explore the effect of ED threshold on throughput.  

\subsection{Coexistence Experiment and Comparison with Analytical Derivation}

The NI 802.11 Labview Application Framework provides functional elements of the Physical (PHY) layer as well as the Medium Access Control (MAC) layer. The module code includes receiver (RX) and transmitter (TX) functionality and elements for channel state handling, slot timing management, and backoff procedure handling. The MAC is implemented on a Field Programmable Gate Array (FPGA) and tightly integrated with the PHY to fulfill the requirements for interframe spacing (such as SIFS, and DIFS), as well as slot timing management to allow frame exchange sequences, such as DATA, ACK and basic DCF for CSMA/CA protocol. 

\begin{figure}[t]
\setlength{\belowcaptionskip}{-0.1in}
\centerline{\includegraphics[width=3.3in]{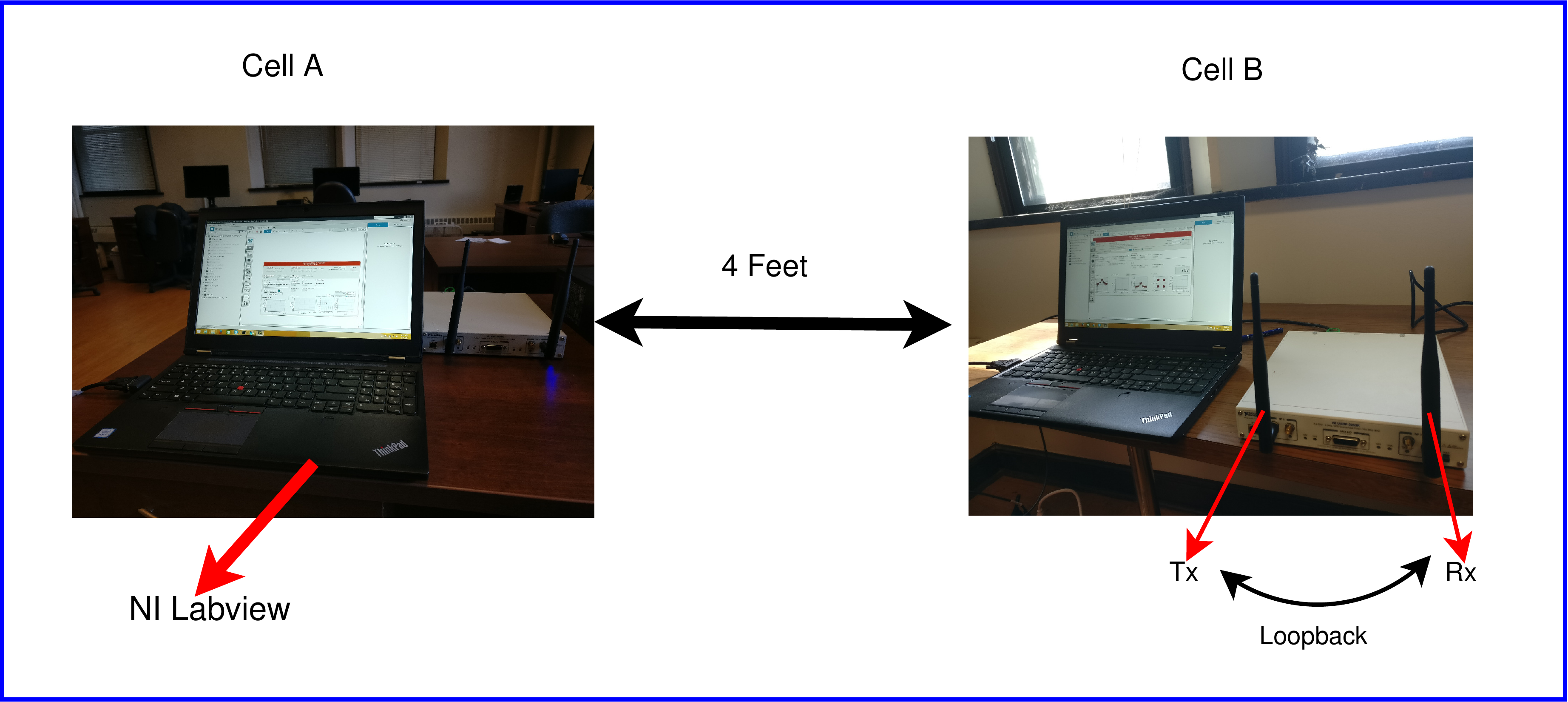}}
 \caption{The Wi-Fi only and LTE-LAA/Wi-Fi coexistence experimental setup.}
 \label{fig: expset}
\end{figure}

Our experimental platform uses National Instruments USRP 2953 R software defined radios (SDRs) as illustrated in Fig.~\ref{fig: expset}. The NI boards used in our set-up can be configured either with two Wi-Fi APs or two LTE-LAA NI eNBs. As a benchmark comparison, we study two NI Wi-Fi APs (i.e., both Cell A and Cell B in Fig.~\ref{fig: expset} are Wi-Fi) which are co-channel (on Channel 161) and contending to transmit downlink data. Full buffer is assumed for both Wi-Fi APs. The hardware requirements consist of two computers which have at least 8 GB RAM (Installed Memory), 64-bit operating system, x64-based processor, Intel(R) Core i7, CPU clock 2.60GHz. The setup also contains 4 antennas covering the 5 GHz ISM band. To study the impact of collisions in the medium we further increased the number of Wi-Fi nodes to 4 (\emph{i.e.,} 2 NI Wi-Fi and 2 Netgear Wi-Fi APs) \& 6 (\emph{i.e.,} 2 NI Wi-Fi and 4 Netgear Wi-Fi APs). These APs are all configured through laptops (via LAN connection) with the parameters mentioned in Table~\ref{table: WiFipar}.

\begin{table}
\caption{Wi-Fi and LTE-LAA parameters.}
\label{table: WiFipar}
\begin{center}
\begin{tabular}{|c|c|}
\hline 
Parameter & value \\
\hline
\hline
 PhyH & 20 $\mu s$ \\
 \hline
 MACH & (34 bytes)$/r_w$ $\mu s$ \\
 \hline
 $r_0$ & 6 Mbps \\
 \hline
 ACK & (14 bytes)$/r_0$ $\mu s$ \\
 \hline
 $\delta$ & 0.1 $\mu s$ \\
 \hline
 $\sigma$ & 9 $\mu s$ \\
 \hline
 $e_l$ & 1 \\
 \hline
 DIFS & 34 $\mu s$ \\
  \hline
 SIFS & 16 $\mu s$ \\
  \hline
 $N_B$ & 2048 bytes \\
  \hline
 Psize & $N_B/r_w$ $\mu s$ \\
 \hline
 $D_{LTE}$ & 0.5 $ms$ \\
 \hline
\end{tabular}
\end{center}
\end{table}

For the coexistence of one Wi-Fi and one LTE-LAA, one of the SDRs is configured to operate as an LTE-LAA eNB and the other is configured as a Wi-Fi AP (i.e., Cell A is LTE-LAA and Cell B is Wi-Fi in Fig.~\ref{fig: expset}). Similarly, for the other two coexistence experiments, two Wi-Fi \& two LTE-LAA and four Wi-Fi \& two LTE-LAA are deployed. The NI LTE-LAA, NI Wi-Fi AP, and Netgear Wi-Fi APs are provisioned to transmit in the DL to their clients (one client per AP/eNB) on the same channel, Channel 161. Wi-Fi uses the 802.11a standard with 20 MHz bandwidth and LTE-LAA uses 100 RBs to cover the total 20 MHz bandwidth and 1 OFDM symbol in a subframe is assigned as the PDCCH symbol. The SDR transmission characteristics along with other experimental parameters under study are summarized in Table \ref{table: NIlab}.

\begin{table}
\caption{NI labview experimental setup parameters.}
\label{table: NIlab}
\begin{center}
\begin{tabular}{|c|c|c|c|}
\hline 
Parameters & NI Experiment \\ 
\hline
\hline
Transmission Power & 23 dBm \\ 
 \hline
Operating Channel & 161 \\ 
 \hline
Operating Frequency & 5.580 GHz \\
 \hline
RF Transmission & Loopback \\
\hline
LTE-LAA Transmission Channel & PDSCH, PDCCH \\
\hline 
Data Traffic & Full buffer \\
 \hline
\end{tabular}
\end{center}
\vspace{-0.2in}
\end{table}

\textbf{Scenario for exploring the throughput performance:} We change the number of nodes in each technology along with their data rates, TXOPs, and channel access parameters to investigate the throughput performance. Three different experiments are performed to compare throughput performance with the theoretical results calculated from the analytical models:
\begin{itemize}
\item
Case 1 (Wi-Fi only): This is the baseline for comparison. The contending nodes are configured to be Wi-Fi APs on the same channel, with 1 client per AP. The experiment is performed for video data transmission with $W_0=16$, $m=6$, and packet length of $N_B=2048$ bytes. 
\item
Case 2 (Coexistence, LTE-LAA Class 1): One NI SDR or Netgear is configured to be the Wi-Fi AP and the other NI SDR is configured to be an LTE-LAA eNB, with 1 client per AP/eNB. The LTE-LAA eNB uses access priority class 1 \cite{ETSILAA} with $W'_0=4$, $m'=1$, and TXOP $=2 ms$ and the Wi-Fi AP uses data transmission with $W_0=4$, $m=1$, and packet length of $N_B=2048$ bytes.
\item
Case 3 (Coexistence, LTE-LAA Class 3): NI SDRs or Netgear  is configured to be the Wi-Fi AP and the other NI SDR as the LTE-LAA eNB, 1 client per AP/eNB. The LTE-LAA eNB uses access priority class 3 \cite{ETSILAA} with $W'_0=16$, $m'=2$, and TXOP $=8 ms$ and the Wi-Fi AP uses video transmission with channel access parameters of $W_0=16$, $m=2$, and packet length of $N_B=2048$ bytes. 
\end{itemize}

\begin{table}
\caption{NI experiment compared with the theoretical derivation for three cases for 2 Wi-Fi node for Wi-Fi only scenario (Wi-Fi Aggregate) and one Wi-Fi / one LTE-LAA for the coexistence scenario.}
\label{table: CompExperiment2}
\begin{center}
\begin{tabular}{|c|c|c|c|}
\hline 
System & NI Experiment & Theoretical Modeling \\
\hline
\hline
\multicolumn{3}{|c|}{Wi-Fi rate $r_w=9$ Mbps and LTE-LAA rate $r_l=7.8$ Mbps} \\
\hline
Case 1 (Wi-Fi Aggregate) & 8.0 & 7.77 \\
 \hline
Case 2 (Wi-Fi \& LAA) & 3.40 and 3.91 & 3.25 and 3.01 \\
 \hline
Case 3 (Wi-Fi \& LAA) & 1.80 and 5.22 & 1.49 and 5.26 \\
\hline
\hline
\multicolumn{3}{|c|}{Wi-Fi rate $r_w=18$ Mbps and LTE-LAA rate $r_l=15.6$ Mbps} \\
\hline
Case 1 (Wi-Fi Aggregate) & 15.0 & 14.62\\
\hline
Case 2 (Wi-Fi \& LAA) & 4.20 and 8.52 & 4.04 and 7.24 \\
\hline
Case 3 (Wi-Fi \& LAA) & 1.69 and 11.09 & 1.63 and 11.51 \\
\hline
\hline
\multicolumn{3}{|c|}{Wi-Fi rate $r_w=54$ Mbps and LTE-LAA rate $r_l=70.2$ Mbps} \\
\hline
Case 1 (Wi-Fi Aggregate) & 35.40 & 34.38\\
\hline
Case 2 (Wi-Fi \& LAA) & 5.20 and 43.10 & 4.71 and 37.90 \\
\hline
Case 3 (Wi-Fi \& LAA) & 1.40 and 57.80 & 1.73 and 55.18 \\
\hline
\end{tabular}
\end{center}
\vspace{-0.2in}
\end{table}

We perform the experiment with three different data rates for Wi-Fi: 9 Mbps (BPSK, coding rate $= 0.5$), 18 Mbps (QPSK, code rate $= 0.75$), and 54 Mbps (64QAM, code rate $= 0.75$). LTE-LAA uses all 100 RBs which yields three data rates of 7.8 Mbps (QPSK, code rate $=0.25$), 15.6 Mbps (QPSK, code rate $=0.5$), and 70.2 Mbps (64QAM, code rate $=0.75$). In the NI Labview implementation, the retransmission stage ($i$) resets to zero after $i$ exceeds $m'$ and the LTE-LAA $D_{LTE}$ is equal to DIFS, so we set $e_l=0$ and DIFS $=D_{LTE}$ in the analytical model to compare with the NI Labview data.

The results for two contending nodes (two Wi-Fi APs for Wi-Fi only and one Wi-Fi AP/one LTE-LAA eNB for coexistence scenario) are shown in Table~\ref{table: CompExperiment2} (all values in the table are Mbps), where the theoretical throughput of the Wi-Fi only network is from Eq. (13) in \cite{Bianchi} using the $\tau_w$ (\ref{eq: tauw}) calculated in this paper and the theoretical throughput of the coexistence network is from (\ref{eq: tputw}) and (\ref{eq: tputl}). As can be seen in Table~\ref{table: CompExperiment2}, the trend of aggregate experimental throughputs in the Wi-Fi only network (Case 1) are similar to the theoretical results for different data rates. In Case 2 and 3, the measured experimental throughputs for the coexistence of Wi-Fi and LTE-LAA show a similar trend to theoretical results. The experimental throughput values are usually larger than the theoretical values because in the theoretical modeling we assumed that any partial overlap of the transmitted frames from different nodes results in a collision; however, this may not happen in practice since the receiver node may be able to decode packets which are partially overlapped (particularly when the overlapped packet has lower interference power) \cite{Interference1, Interference2}. 
Especially, if we compare the results of Case 2 and Case 3 in Table~\ref{table: CompExperiment2} with fixed $r_w=9$ Mbps and $r_l=7.8$ Mbps, we see that the gap between the theoretical and experimental throughput of the LTE-LAA in Case 2 is larger than Case 3. We believe that this is due to  a smaller contention window and TXOP in Case 2, which results in more access to the channel. As channel access increases, the number of frames overlapping with other frames increases. All overlaps are considered as collision in the theory, but in the experiment only part of these overlapped frames lead to collision; so, the gap increases. 
As the data rate increases, the relative gap between theory and experiment decreases for different sets of Wi-Fi and LTE-LAA parameters. Higher data rates (higher code rate and modulation scheme) are more sensitive to the interference, therefore the overlap of the packets leads more to collision in reality, which brings it closer to the theory. 

The total theoretical throughput in Case 2 for the coexistence scenario is 6.26 Mbps (for $r_w=9$ Mbps and $r_l=7.8$ Mbps) which is smaller than the aggregate theoretical throughput of 7.78 Mbps in Wi-Fi only. In Case 3, the total theoretical coexistence throughput is 6.75 Mbps (for $r_w=9$ Mbps and $r_l=7.8$ Mbps) which is again smaller than the theoretical Wi-Fi only throughput but by a lower margin as compared with Case 2. The reasons behind the smaller throughput of the coexistence network compared to the Wi-Fi only network is discussed in the next sub-section.

\begin{table}
\caption{NI experiment compared with the theoretical derivation for three cases for 4 Wi-Fi node for Wi-Fi aggregate throughput and 2 Wi-Fi / 2 LTE-LAA for the coexistence scenario.}
\label{table: CompExperiment4}
\begin{center}
\begin{tabular}{|c|c|c|c|}
\hline 
System & NI Experiment & Theoretical Modeling \\
\hline
\hline
\multicolumn{3}{|c|}{Wi-Fi rate $r_w=9$ Mbps and LTE-LAA rate $r_l=7.8$ Mbps} \\
\hline
Case 1 (Wi-Fi Aggregate) & 7.89 & 7.24 \\
\hline
Case 2 (Wi-Fi \& LAA) & 2.07 and 2.30 & 2.18 and 1.94 \\
\hline
Case 3 (Wi-Fi \& LAA) & 1.31 and 3.82 & 1.34 and 4.72 \\
\hline
\hline
\multicolumn{3}{|c|}{Wi-Fi rate $r_w=18$ Mbps and LTE-LAA rate $r_l=15.6$ Mbps} \\
\hline
Case 1 (Wi-Fi Aggregate) & 13.90 & 13.73\\
\hline
Case 2 (Wi-Fi \& LAA) & 2.82 and 4.94 & 2.68 and 4.66 \\
\hline
Case 3 (Wi-Fi \& LAA) & 1.62 and 9.98 & 1.46 and 10.24 \\
\hline
\hline
\multicolumn{3}{|c|}{Wi-Fi rate $r_w=54$ Mbps and LTE-LAA rate $r_l=70.2$ Mbps} \\
\hline
Case 1 (Wi-Fi Aggregate) & 34.78 & 34.07\\
\hline
Case 2 (Wi-Fi \& LAA) & 3.12 and 25.18 & 2.93 and 23.30 \\
\hline
Case 3 (Wi-Fi \& LAA) & 1.98 and 51.20 & 1.54 and 48.98 \\
\hline
\end{tabular}
\end{center}
\end{table}

\begin{table}
\caption{NI experiment compared with the theoretical derivation for three cases for 6 Wi-Fi node for Wi-Fi aggregate throughput and 4 Wi-Fi / 2 LTE-LAA for the coexistence scenario.}
\label{table: CompExperiment6}
\begin{center}
\begin{tabular}{|c|c|c|c|}
\hline 
System & NI Experiment & Theoretical Modeling \\
\hline
\hline
\multicolumn{3}{|c|}{Wi-Fi rate $r_w=9$ Mbps and LTE-LAA rate $r_l=7.8$ Mbps} \\
\hline
Case 1 (Wi-Fi Aggregate) & 7.16 & 6.90 \\
\hline
Case 2 (Wi-Fi \& LAA) & 2.0 and 1.03 & 1.93 and 0.85 \\
\hline
Case 3 (Wi-Fi \& LAA) & 1.86 and 3.10 & 2.01 and 3.56 \\
\hline
\hline
\multicolumn{3}{|c|}{Wi-Fi rate $r_w=18$ Mbps and LTE-LAA rate $r_l=15.6$ Mbps} \\
\hline
Case 1 (Wi-Fi Aggregate) & 13.17 & 13.12\\
\hline
Case 2 (Wi-Fi \& LAA) & 2.67 and 2.43 & 2.42 and 2.14 \\
\hline
Case 3 (Wi-Fi \& LAA) & 2.66 and 7.76 & 2.31 and 8.19 \\
\hline
\hline
\multicolumn{3}{|c|}{Wi-Fi rate $r_w=54$ Mbps and LTE-LAA rate $r_l=70.2$ Mbps} \\
\hline
Case 1 (Wi-Fi Aggregate) & 33.19 & 32.85\\
\hline
Case 2 (Wi-Fi \& LAA) & 3.00 and 14.06 & 2.91 and 11.55 \\
\hline
Case 3 (Wi-Fi \& LAA) & 2.98 and 43.11 & 2.57 and 40.99 \\
\hline
\end{tabular}
\end{center}
\end{table}

Similarly, the scenario is extended for 4 contending nodes (4 Wi-Fi APs for Wi-Fi only and 2 Wi-Fi APs/2 LTE-LAA eNBs for coexistence scenario) and 6 contending nodes (6 Wi-Fi APs for Wi-Fi only and 4 Wi-Fi APs/2 LTE-LAA eNBs for coexistence scenario). The results are presented in Table~\ref{table: CompExperiment4} and \ref{table: CompExperiment6}. As the number of Wi-Fi and LTE-LAA contending nodes increases, the trend of the throughput in experimental and theoretical results are the same for both aggregate throughput in Wi-Fi only network and Wi-Fi/LTE-LAA aggregate throughput in the coexistence scenario. Also, by increasing the number of nodes, we observe the gap between the experiment and theory due to the same aforementioned reasons for the gaps in Table~\ref{table: CompExperiment2}. 

As the number of Wi-Fi and LTE-LAA nodes increase, the collision events increase resulting in lower throughput both in Wi-Fi only network and coexistence network. By increasing the number of nodes, the coexistence network achieves a lower throughput then the Wi-Fi only network (e.g. in Table~\ref{table: CompExperiment4}; for $r_w=9$ Mbps and $r_l=7.8$ Mbps, coexistence system achieves the total throughput of 4.12 Mbps in Case 2 and 6.06 Mbps in Case 3 which is smaller than the Wi-Fi only network throughput of 7.24 Mbps), because the LTE-LAA TXOP is larger than Wi-Fi, any collision to LTE-LAA results in greater loss of airtime for transmission.

\subsection{Numerical Results from Analytical Derivation}

Having validated our analysis with experiments in the previous section, we now evaluate the coexistence performance of Wi-Fi and LTE-LAA for other scenarios via numerical evaluation of results in Sec. IV, for the parameters listed in Table~\ref{table: WiFipar}. For LTE-LAA LBT the class 2 and 4 parameters from Table~\ref{table: LAAclass} are used. The per user throughput in each network is calculated by dividing the total throughput over the number of nodes in that network (i.e. Wi-Fi per user throughput in a coexistence scenario is $\frac{Tput_w}{n_w}$). 

\begin{figure}[t]
\setlength{\belowcaptionskip}{-0.1in}
\centerline{\includegraphics[width=3.3in]{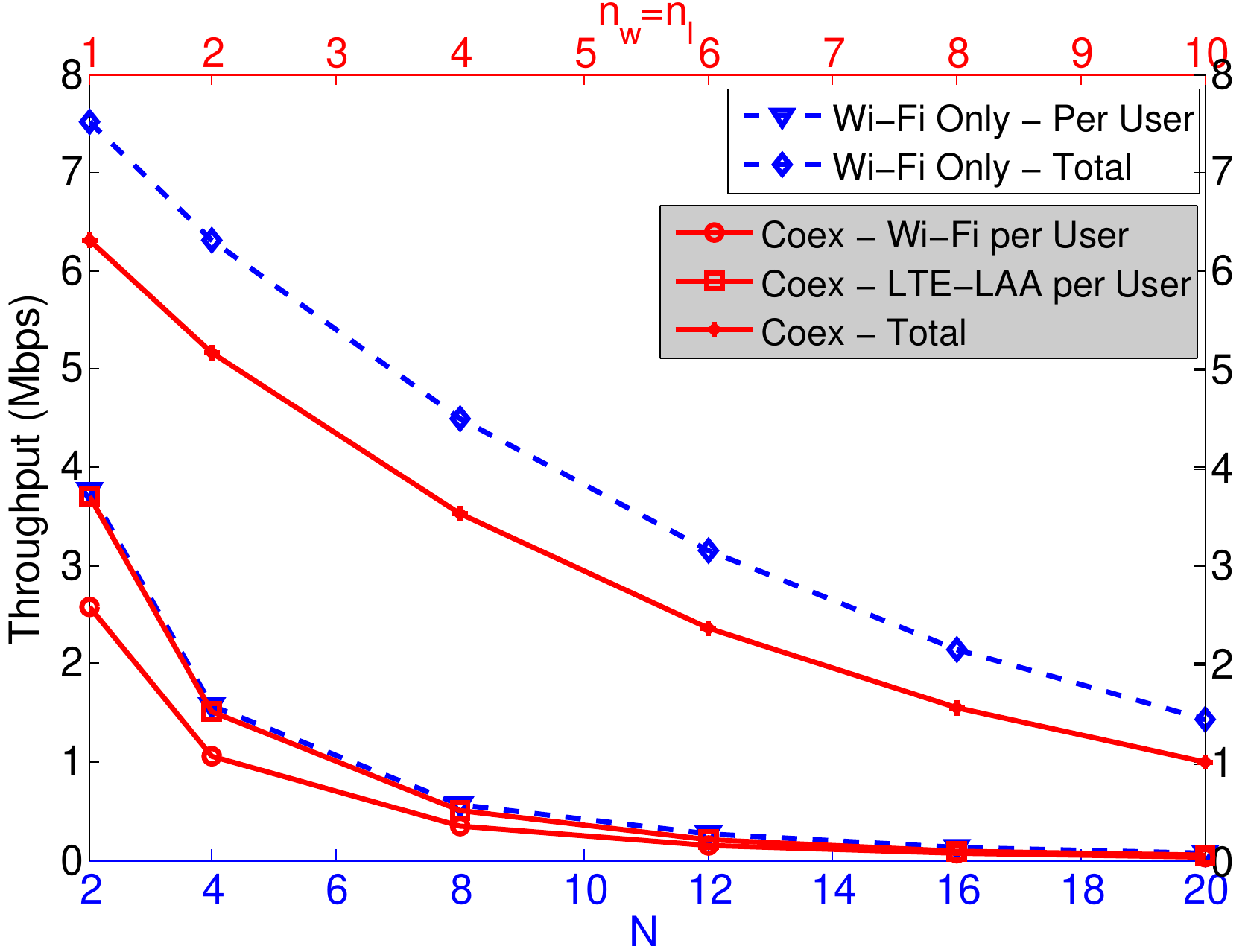}}
 \caption{Throughput of the coexistence system (considering priority class 2 for LTE-LAA) compared with the Wi-Fi Only, $W_0=W'_0=8$, $m=m'=1$. Bottom x-axis (blue) is the number of Wi-Fi APs in Wi-Fi only network and top x-axis (red) is the number of Wi-Fi APs and LTE-LAA eNBs in coexistence network.}
 \label{fig: sim0}
\end{figure}

\textbf{Exploring difference in TXOP of Wi-Fi and LTE-LAA:} In Fig.~\ref{fig: sim0} we show the results for the Wi-Fi only system with $N$ APs and the coexistence system with $n_w=n_l=N/2$, $W_0=W'_0=8$, $m=m'=1$, LTE-LAA TXOP  $=3~ms$, $r_w=9$ Mbps, and $r_w=8.4$ Mbps. The other parameters are listed in Table~\ref{table: WiFipar}. The number of Wi-Fi APs $N$ in the Wi-Fi only network is equally divided among the number of Wi-Fi APs and LTE-LAA eNBs in the coexistence network, i.e. $n_w=n_l=N/2$. We observe that the total throughput of Wi-Fi and LTE-LAA coexistence is lower than the total throughput of the Wi-Fi only network. In part this is due to LTE-LAA channel access obeying strict slot boundaries for transmission (i.e. LTE-LAA transmits a reservation signal to keep the channel until the start of the next slot) that wastes some fraction of channel use and hence degrades the coexistence throughput. Further, since the transmission duration of Wi-Fi is smaller than the LTE-LAA TXOP, the per user throughput of Wi-Fi in coexistence is lower than the LTE-LAA.

\begin{figure}[t]
\setlength{\belowcaptionskip}{-0.1in}
\centerline{\includegraphics[width=3.3in]{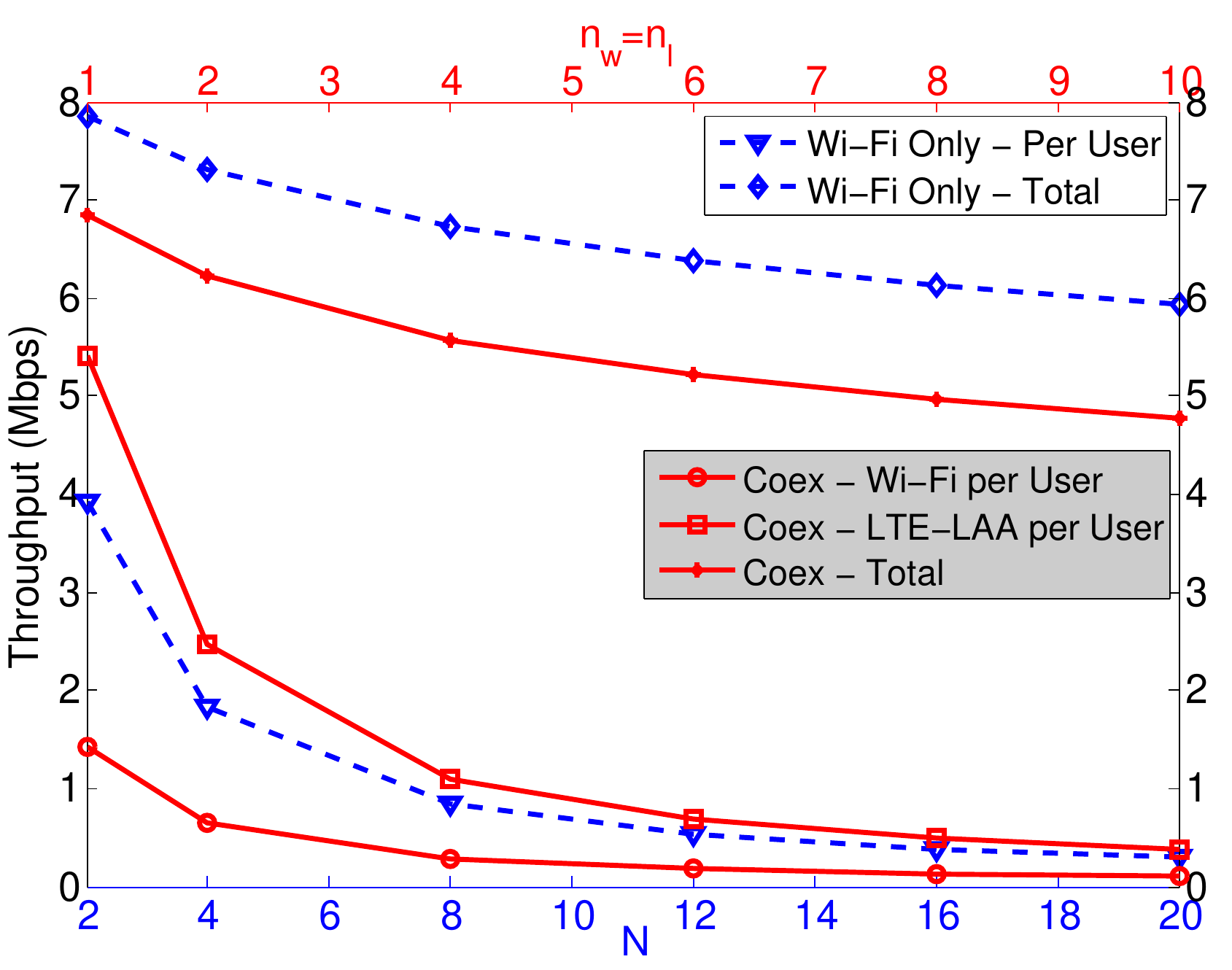}}
 \caption{Throughput of the coexistence system (considering priority class 4 for LTE-LAA) compared with the Wi-Fi Only considering $W_0=W'_0=16$, $m=m'=6$.}
 \label{fig: sim1}
\end{figure}

\textbf{Exploring the effect of TXOP and contention parameters:} In Fig.~\ref{fig: sim1}, the Wi-Fi only system with $N$ APs is compared against the coexistence system with $n_w=n_l=N/2$, $W_0=W'_0=16$, $m=m'=6$, LTE-LAA TXOP $=8~ms$, $r_w=9$ Mbps, and $r_w=8.4$ Mbps; other parameters are listed in Table~\ref{table: WiFipar}. The total throughput of Wi-Fi and LTE-LAA is lower than the total throughput of Wi-Fi only network. In addition to the wasted channel opportunity for transmission due to the reservation signal, the transmission duration of LTE-LAA is more than Wi-Fi which causes LTE-LAA to have higher per-user throughput than the Wi-Fi only. The larger difference between coexistence and Wi-Fi only network throughput curves compared with Fig.~\ref{fig: sim0} results because the LTE-LAA gets a higher opportunity for access and has a lower data rate compared with Wi-Fi. 

\begin{figure}[t]
\setlength{\belowcaptionskip}{-0.1in}
\centerline{\includegraphics[width=3.3in]{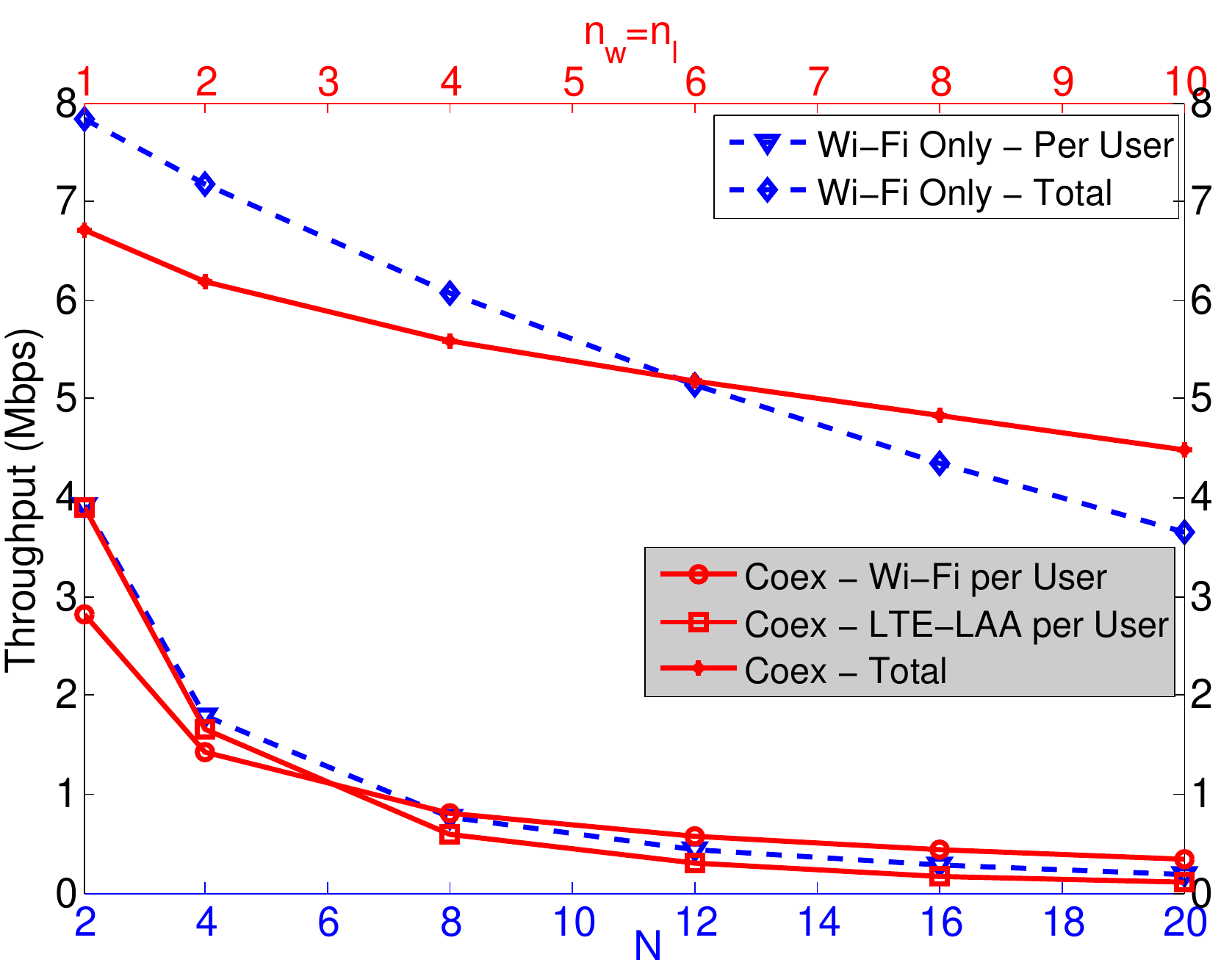}}
 \caption{Throughput of the coexistence system with $W_0=16$, $m=1$ and LTE-LAA with $W'_0=16$, $m'=6$.}
 \label{fig: sim3}
\end{figure}

\textbf{Exploring the difference in contention parameters of Wi-Fi and LTE-LAA:} In Fig.~\ref{fig: sim3}, the throughput performance of Wi-Fi APs with $W_0=16$, $m=1$, $r_w=9$ Mbps, and LTE-LAA eNB with $W'_0=16$, $m'=6$, TXOP $=3~ms$, and $r_w=8.4$ Mbps is shown. The other parameters are listed in Table~\ref{table: WiFipar}. The Wi-Fi per user throughput in coexistence for $N\ge8$ is higher than the Wi-Fi only per user throughput, which indicates that fair coexistence based on the 3GPP definition \cite{3GPP_TR} is achieved. The other observation is that the throughput of the coexistence network compared with the Wi-Fi only network achieves a) smaller total throughput for the fewer number of stations ($N<12$) but b) higher throughput at the number of stations $N\ge12$. This is due to the smaller maximum retransmission stage of Wi-Fi (i.e., $m$) in which the Wi-Fi stations access the channel more frequently, as well as the fewer number of Wi-Fi stations in coexistence network, i.e. $n_w=\frac{N}{2}$, which causes fewer collision. This implies that in a specific network setup, the total throughput of coexistence network could be higher or lower than Wi-Fi only network depending on the number of stations.  

\begin{figure}[t]
\setlength{\belowcaptionskip}{-0.1in}
\centerline{\includegraphics[width=3.3in]{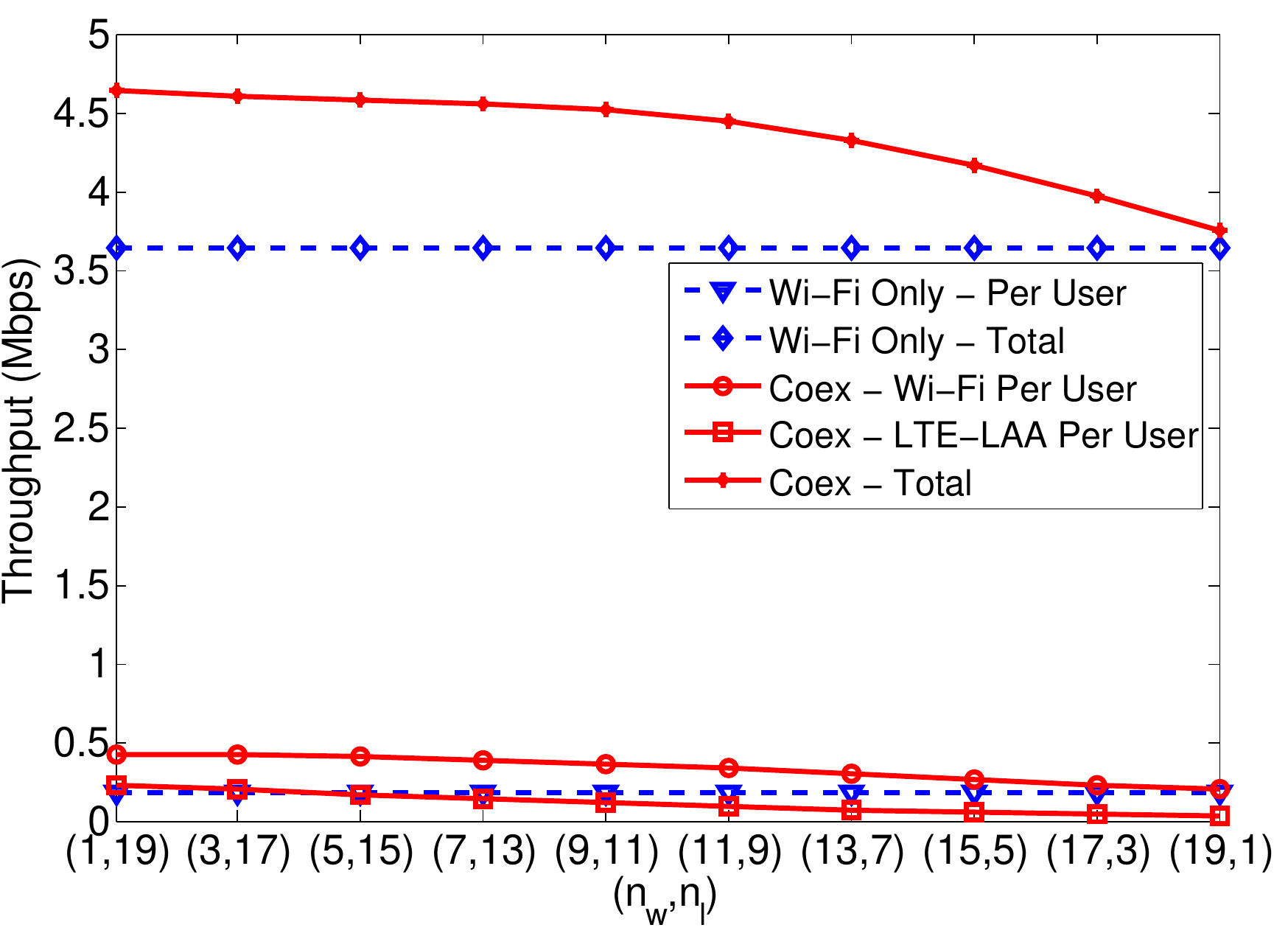}}
 \caption{Throughput of the coexistence system with $W_0=16$, $m=1$ and LTE-LAA with $W'_0=16$, $m'=6$.}
 \label{fig: sim4}
\end{figure}

\textbf{Asymmetric number of Wi-Fi and LTE-LAA nodes:} The setup in Fig.~\ref{fig: sim4} exactly follows that in Fig.~\ref{fig: sim3}, but the total number of stations is fixed at 20, i.e., $N=n_w+n_l=20$, while the number of LTE-LAA eNBs and Wi-Fi APs in the coexistence network are varying. The goal is to investigate the effect of the asymmetric number of Wi-Fi and LTE-LAA stations in coexistence network. The maximum throughput of the coexistence system occurs at $n_w=1, n_l=19$ which indicates that a larger number of LTE-LAA eNBs delivers a higher portion of total throughput in this scenario. Moreover, the throughput of coexistence network for any combination of Wi-Fi and LTE-LAA is higher than the Wi-Fi only system. By increasing the number of Wi-Fi and decreasing the number of LTE-LAA nodes, the throughput of both systems is seen to decrease. For the total range of the number of stations, the Wi-Fi {\em per user} throughput in coexistence network is higher than the Wi-Fi per user throughput in Wi-Fi only system, which illustrates that the 3GPP fairness is achieved regardless of the number of stations for this (per user throughput) metric.  

\begin{figure}[t]
\setlength{\belowcaptionskip}{-0.1in}
\centerline{\includegraphics[width=3.3in]{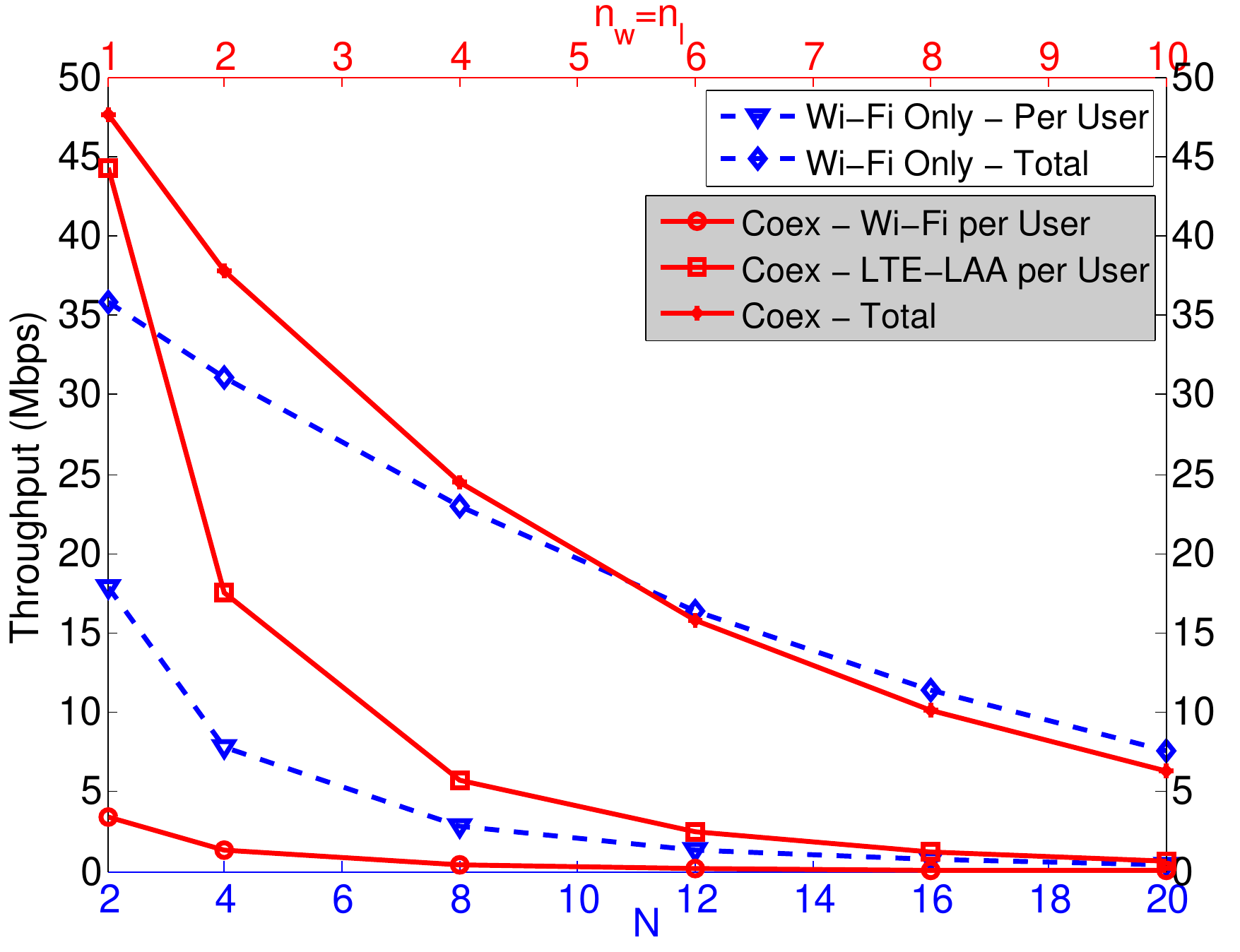}}
 \caption{Throughput of the coexistence system with $W_0=8$, $m=1$, $W'_0=8$, $m'=1$ and the higher data rates for Wi-Fi and LTE-LAA $r_w=54$ Mbps and $r_l=75.8$ Mbps.}
 \label{fig: sim5}
\end{figure}

\begin{figure}[t]
\setlength{\belowcaptionskip}{-0.1in}
\centerline{\includegraphics[width=3.3in]{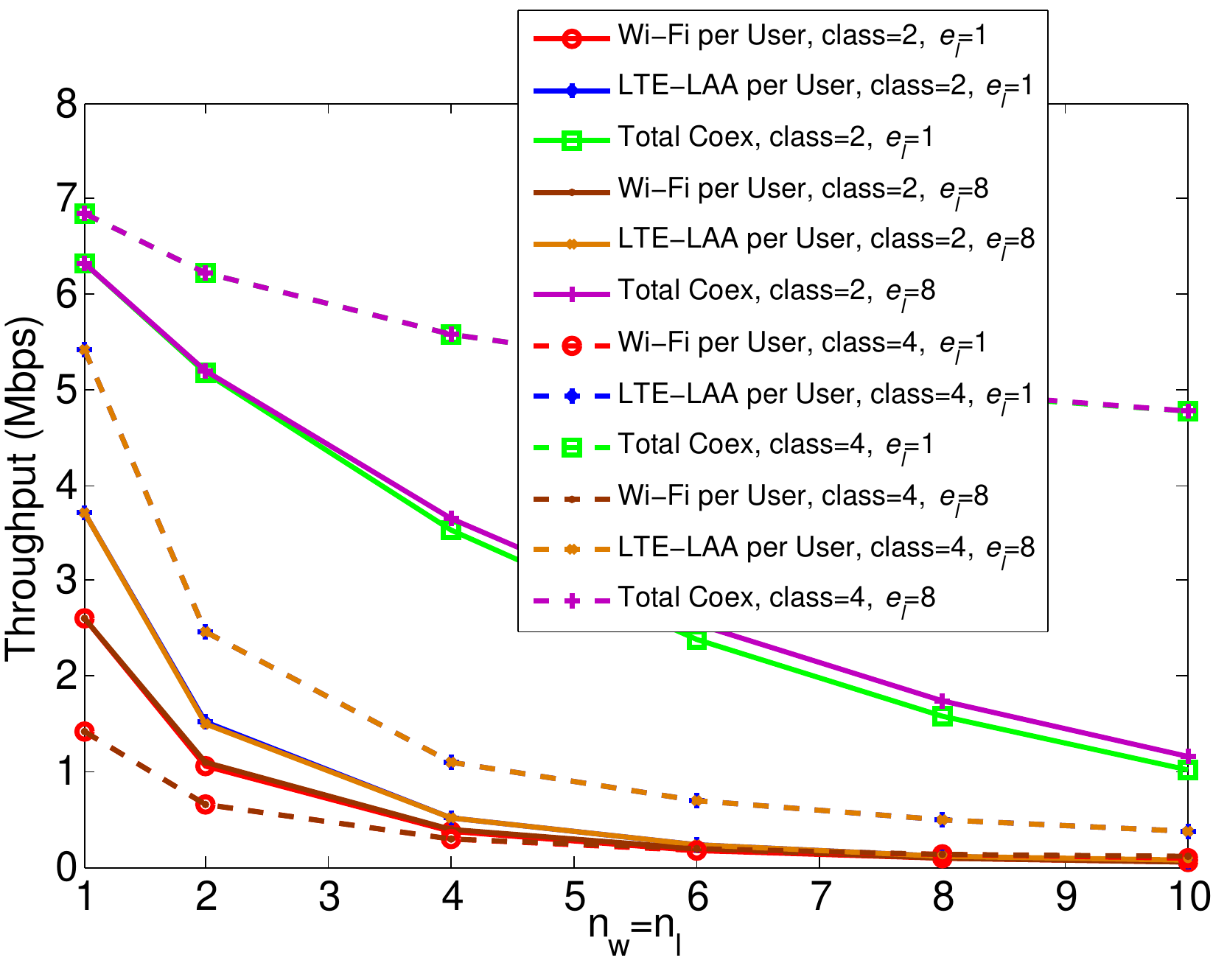}}
 \caption{Throughput versus different values of $e_l$ in LTE-LAA.}
 \label{fig: sim6}
\end{figure}

\textbf{Exploring the higher data rate for Wi-Fi and LTE-LAA:} The throughput performance of the coexistence system with the parameters $W_0=W'_0=8$, $m=m'=1$, LTE-LAA TXOP $=3~ms$, and higher data rates for Wi-Fi and LTE-LAA, $r_w=54$ Mbps and $r_w=75.8$ Mbps, is illustrated in Fig.~\ref{fig: sim5}. The other parameters are listed in Table~\ref{table: WiFipar}. The Wi-Fi per user throughput in coexistence is lower than the LTE-LAA per user throughput because the Wi-Fi frame airtime is smaller at higher Wi-Fi data rates while the LTE-LAA airtime (TXOP) is fixed. Smaller airtime of Wi-Fi, considering the same channel access parameters as LTE-LAA, leads to higher utilization of the channel by LTE-LAA. This also results in lower Wi-Fi per user throughput in coexistence compared with per user throughput of Wi-Fi only network, because in Wi-Fi only network the airtime and channel access parameters are the same for all users while in coexistence scenario, the LTE-LAA has a higher airtime. In Fig.~\ref{fig: sim6}, the effect of changing the $e_l$ (retry limit after reaching to $m'$) the  in LTE-LAA is investigated for different scenarios; the curves for class 2 follow the parameters of Fig.~\ref{fig: sim0} and class 4 follow the Fig.~\ref{fig: sim1}. For class 2 where $W'_0=W_0=8$ and $m'=m=1$, the total throughput as well as the per user throughput of Wi-Fi and LTE-LAA increases noticeably with retry limit, but much less for class 4 with $W'_0=W_0=16$ and $m'=m=6$. 

\subsection{Effect of ED threshold on Coexistence throughput}

\begin{figure}[t]
\setlength{\belowcaptionskip}{-0.1in}
\centerline{\includegraphics[width=3.3in]{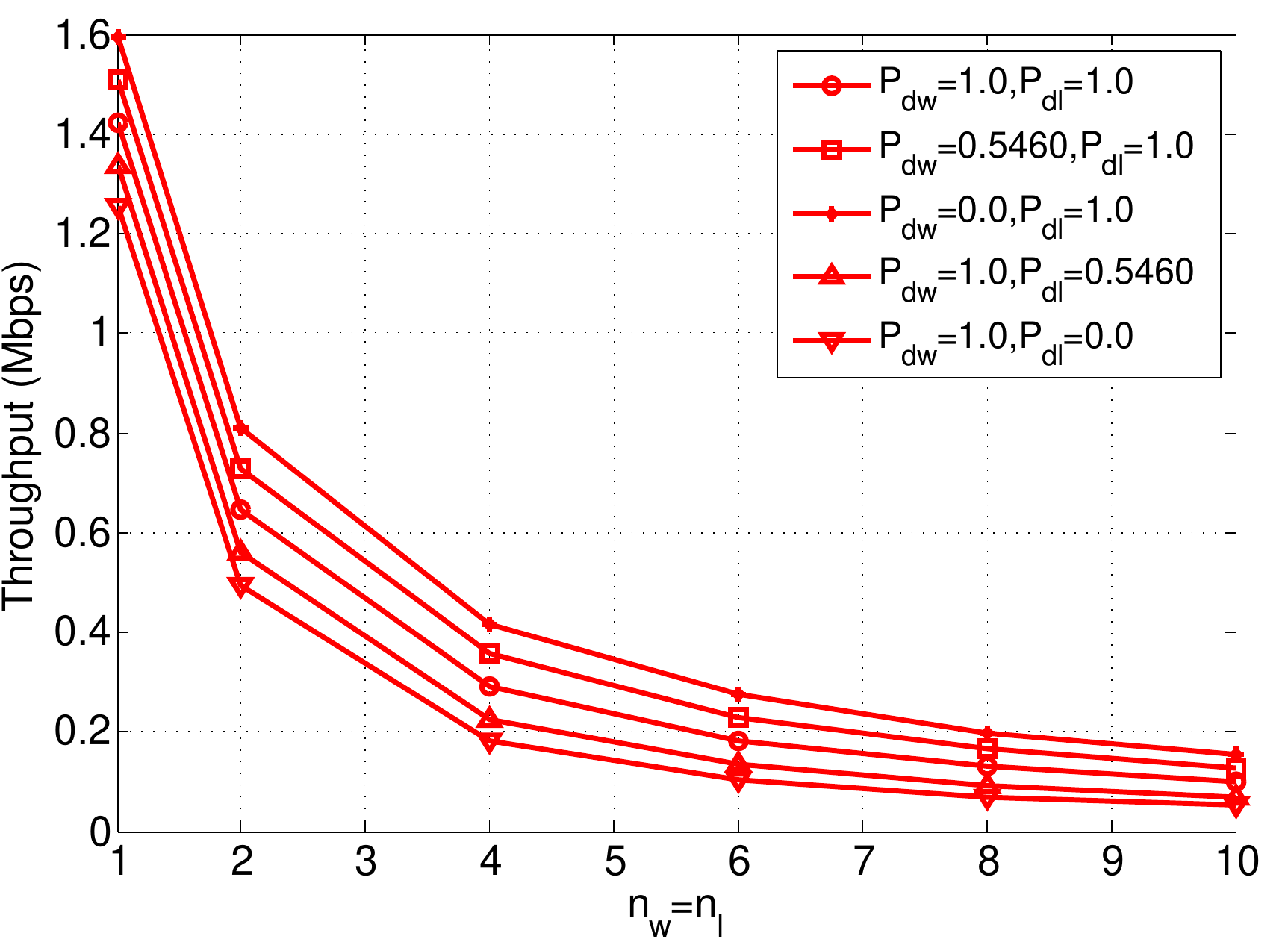}}
 \caption{Throughput performance of Wi-Fi through changing the detection probability of Wi-Fi and LTE-LAA.}
 \label{fig: ED0}
\end{figure}

In Fig.~\ref{fig: ED0}, numerical results following the derivations in Sec. V illustrates the effect of varying  ED threshold on the throughput of Wi-Fi in coexistence network with $W_0=W'_0=16$, $m=m'=6$, LTE-LAA TXOP $=8~ms$, and Wi-Fi packet length 2048 bytes. Table~\ref{table: energy} shows the ED probability calculated using (\ref{eq: Pd}) for different ED thresholds during the DIFS period with 680 samples, and other parameters as listed in Table~\ref{table: WiFipar}. By increasing the detection probability of Wi-Fi ($P_{dw}$) while keeping the detection probability of LTE-LAA ($P_{dl}$) (nearly) equal to 1, the Wi-Fi throughput decreases;  similarly by increasing the detection probability of LTE-LAA (keeping the detection probability of Wi-Fi nearly equal to 1) the Wi-Fi throughput increases. By decreasing the ED threshold in Wi-Fi, the Wi-Fi nodes detect  LTE-LAA stations at lower transmit power and thus defer their transmission (i.e., enter back-off) leading to lower Wi-Fi throughput. On the other hand, by decreasing the LTE-LAA ED threshold, the LTE-LAA detects more Wi-Fi stations with lower signal power and defers the transmission, so Wi-Fi stations have more opportunity for transmission which increases the throughput. 

\begin{table}[]
\centering
\caption{Energy Detector parameters and $P_d$.}
\label{table: energy}
\begin{tabular}{|c|c|c|c|}
\hline
Parameter                   & \multicolumn{3}{c|}{Value} \\ \hline
signal to noise ratio (SNR) & \multicolumn{3}{c|}{22 dB}      \\ \hline
$P_n$                 & \multicolumn{3}{c|}{-94 dBm}      \\ \hline
 ED threshold       &    -62 dBm     &    -72 dBm     &   -82 dBm     \\ \hline
ED detection probability ($P_{dw}$, $P_{dl}$)                &     0.0    &    0.5460     &    1.0    \\ \hline
\end{tabular}
\end{table}

\begin{figure}[t]
\setlength{\belowcaptionskip}{-0.1in}
\centerline{\includegraphics[width=3.3in]{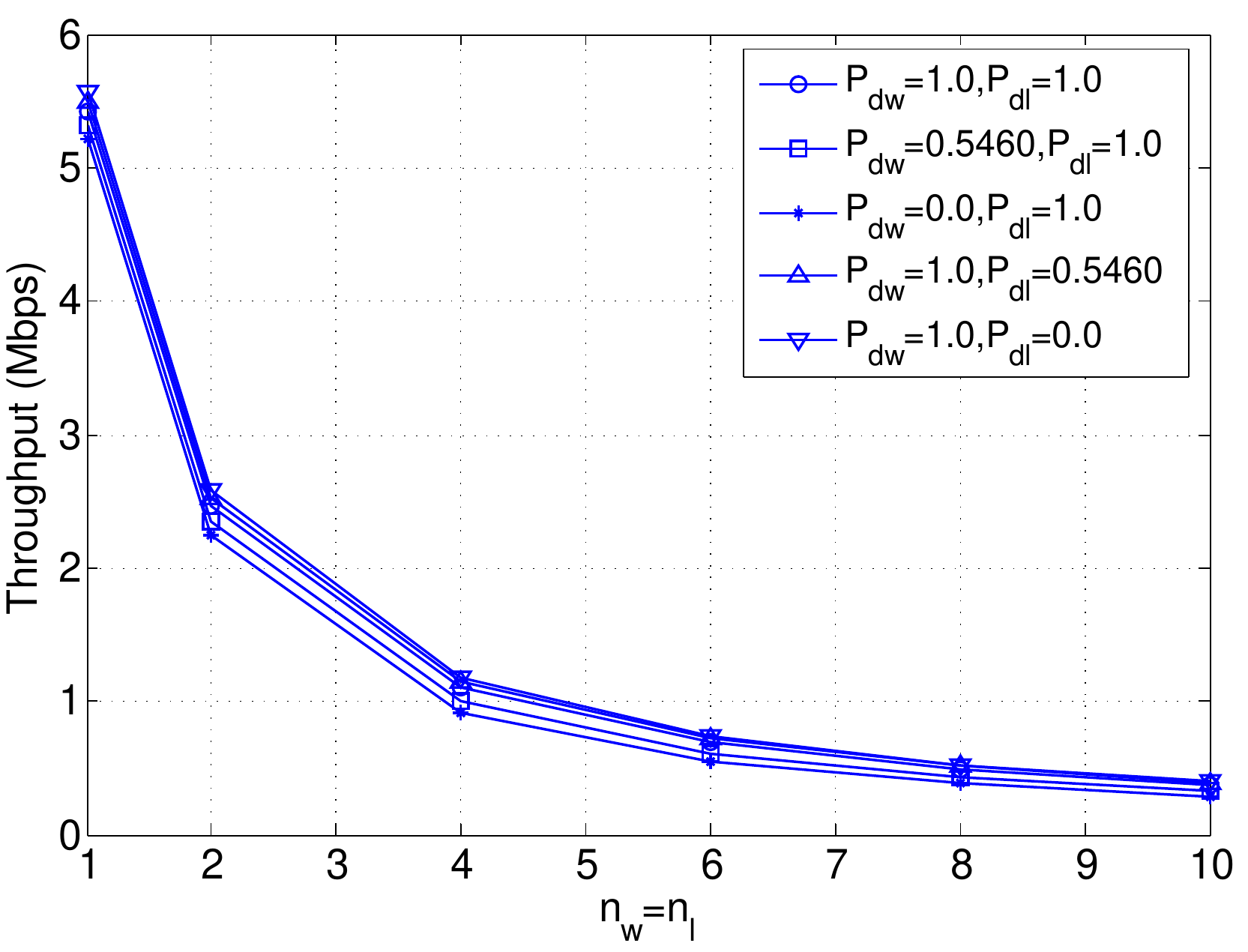}}
 \caption{Throughput performance of LTE-LAA through changing the detection probability of Wi-Fi and LTE-LAA.}
 \label{fig: ED1}
\end{figure}

Fig.~\ref{fig: ED1} shows the effect of changing the detection probability of Wi-Fi and LTE-LAA on the throughput of LTE-LAA in coexistence  similar to Fig.~\ref{fig: ED0}. By increasing the detection probability of Wi-Fi, the LTE-LAA throughput increases, and by increasing the detection probability of LTE-LAA, the LTE-LAA throughput decreases, although impact of changing ED threshold on LTE-LAA throughput is proportionally smaller compared with Wi-Fi. A similar conclusion can be drawn from  Fig.~\ref{fig: ED0}. These results are in line with those presented by NI in \cite{NI_whitepaper}. 

\section{Conclusion}

In this work, we first presented a new model for analyzing the throughput performance for Wi-Fi and LTE-LAA coexistence. We then modified the model to incorporate ED sensing threshold to evaluate the impact of threshold choices on throughput performance. The maximum throughput in a coexistence scenario can be achieved by tuning the ED sensing threshold. To validate the proposed model, we also set up a lab experiment with NI Labview and compared the experimental throughput with the numerical results and showed very good correspondence between experiment and analysis. The throughput performance of a Wi-Fi and LTE-LAA in coexistence system depends on the channel access parameters, TXOP of LTE-LAA, and data rates of Wi-Fi and LTE-LAA. By changing these parameters, the Wi-Fi or LTE-LAA achieves higher per user throughput in coexistence network compared with the per user throughput in Wi-Fi only network. Finally, we note that these results also form the bedrock of a thorough and objective look at coexistence fairness in future, where we expect to show how the CSMA/CA and/or LBT parameters must be tuned to achieve fairness.

\section*{ACKNOWLEDGMENT}
This work was supported by the National Science Foundation (NSF) under grant CNS-1617153. The authors would like to thank Dr. Thomas Henderson, U. Washington for many helpful discussions.

\bibliographystyle{IEEEtran}
\bibliography{WiFiCoex}

\end{document}